\documentclass[english]{scrbook}
\usepackage[scaled=0.9]{helvet}

\pdfoutput=1
\usepackage[T1]{fontenc}
\usepackage[latin9]{inputenc}
\usepackage{wrapfig}
\usepackage{units}
\usepackage{amsthm}
\usepackage{amsmath}
\usepackage{amssymb}
\usepackage{graphicx}
\usepackage{setspace}
\makeatletter

\providecommand{\tabularnewline}{\\}

\numberwithin{equation}{section}
\numberwithin{figure}{section}

\usepackage[labelfont={bf},font=footnotesize,normal]{caption}
\usepackage{xcolor}
\usepackage{sectsty}
\allsectionsfont{\color{darkgray}}  
\usepackage{cite}
\usepackage{hyperref}
\hypersetup{colorlinks=true,citecolor=blue,linkcolor=red}

\makeatother

\usepackage{babel}
\begin{document}

\title{Sodium Magnetic Resonance Imaging:\\
Biomedical Applications %
\thanks{This article is a long version of a review article that will be published
in Journal of Magnetic Resonance Imaging in 2013. %
}\\
~}

\author{Guillaume Madelin\\
~\\
{\normalsize New York University Langone Medical Center}\\
{\normalsize Department of Radiology - Center for Biomedical Imaging}\\
{\normalsize New York, NY 10016, USA}\\
~\\
\includegraphics[width=6cm]{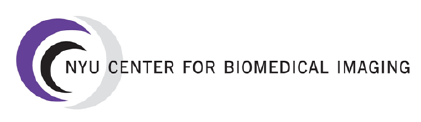}\\
~\\
}

\maketitle
\newpage{}

\chapter*{Abstract}

In this article, we present an up-to-date overview of the potential
biomedical applications of sodium MRI \emph{in vivo}. Sodium MRI is a 
subject of increasing interest in translational research as it can give some
direct and quantitative biochemical information on the tissue viability,
cell integrity and function, and therefore not only help the diagnosis
but also the prognosis of diseases and treatment outcomes. It has
already been applied \emph{in vivo} in most of human tissues, such
as brain for stroke or tumor detection and therapeutic response, in
breast cancer, in articular cartilage, in muscle and in kidney, and
it was shown in some studies that it could provide very useful new
information not available through standard proton MRI. However, this
technique is still very challenging due to the low detectable sodium
signal in biological tissue with MRI and hardware/software limitations
of the clinical scanners. The article is divided in three parts: (1)
the role of sodium in biological tissues, (2) a short review on sodium
magnetic resonance, and (3) a review of some studies on sodium MRI
on different organs/diseases to date.

\newpage{}

\begin{small}

\tableofcontents{}

\end{small}

\newpage{}

\chapter*{Introduction}

\addcontentsline{toc}{chapter}{Introduction} 

Proton magnetic resonance imaging (MRI) can generate more than 30
different kinds of image contrast (and still counting) which can give
the physicians crucial information of the anatomy, and sometimes function,
of the organs under investigation. All organs and most of the known
diseases (if not all) can be studied with this imaging technique in
order to assess their presence and/or degree of possible harm to the
body. However very often, the information from standard MRI cannot
provide direct biochemical information about the health of the tissues,
such as cell integrity and tissue viability, or predict possible outcomes
from different treatments. Sodium MRI could provide some of this complementary
information in a quantitative and non-invasive manner. 

Sodium, symbol $\mathsf{^{23}Na}$, is a quadrupolar nucleus with
spin $\nicefrac{\mathsf{3}}{\mathsf{2}}$ which yields the second
strongest nuclear magnetic resonance (NMR) signal among all nuclei
present in biological tissues, after proton $\mathsf{^{1}H}$. The
NMR sensitivity for sodium is 9.2\% of the proton sensitivity and
the concentration \emph{in} \emph{vivo} is approximately 2000 times
lower than the water proton concentration. As a consequence, sodium
MRI have a average signal-to-noise (SNR) ratio which is around 3000
to 20000 times lower than the proton MRI SNR (depending on organs).
Due to the increase of the available magnetic fields for MRI scanners
(1.5T, 3T, 7T, 9.4T), hardware capabilities such as high gradient
strengths with high slew rates, and new double-tuned radiofrequency
(RF) coils, sodium MRI is now possible in a reasonable time (around
10 to 15 minutes) with a resolution of a few millimeters and has already
been applied \emph{in vivo} in many human organs such as human brain
for tumor and stroke, cartilage, kidneys, heart, muscle or breast
cancer, that will be described in this review article. Sodium MRI
is still a subject of investigation for assessing its medical utility
as a complement to proton MRI or other imaging modalities such as
PET and CT, and determine the new functional/biochemical that it can
provide.

The goal of this review is to describe some potential biomedical applications
of sodium MRI \emph{in vivo} and is divided in three parts: 
\begin{enumerate}
\item \vspace{-3mm}The role of sodium in biological tissues.
\item \vspace{-3mm}A short review on sodium magnetic resonance physics.
\item \vspace{-3mm}An overview of some studies on sodium MRI on different
organs/diseases to date.
\end{enumerate}
\newpage{}

\chapter{Sodium in Biological Tissues}

\section{Homeostasis }

Sodium is a vital component of the human body. It is an important
electrolyte that helps maintain the homeostasis of the organism through
osmoregulation (maintain blood and body fluid volume) and pH regulation
\cite{2008_Burnier}. It is also involved in the cell physiology through
the transmembrane electrochemical gradient, in heart activity, in
the transmission of nerve impulses and muscle contractions through
propagation of action potential in neurons and muscle cells via the
sodium channels. Sodium concentrations are therefore very sensitive
to changes in metabolic state of tissues and integrity of cell membranes.
The intracellular volume fraction is around 80\% of the tissues with
a sodium concentration of 10-15 mM (or mmol/L), and the extracellular
volume fraction (including the vascular compartment) is around 20\%,
with a sodium concentration of 140-150 mM. Cells in healthy tissues
maintain this large sodium concentration gradient between the intracellular
and extracellular compartments across the cell membrane, and any impairment
of energy metabolism or insult to the cell membrane integrity leads
to an increase on intracellular sodium concentration. The sodium influx
and outflux in cells can occur by several routes such as Na$^{+}$
channels, Na$^{+}$/Ca$^{+}$ exchange (NCX), Na$^{+}$/H$^{+}$ exchange
(NHE), Na$^{+}$/bicarbonate (HCO$_{\mathsf{3}}$$^{-}$) cotransporter,
Na$^{+}$/K$^{+}$ /2Cl$^{-}$ cotransporter, Na$^{+}$/Mg$^{+}$
exchange and most importantly through the Na$^{+}$/K$^{+}$-ATPase
\cite{2009_104_3}.

\section{Na$^{+}$/K$^{+}$-ATPase}

\begin{wrapfigure}{o}[30mm]{80mm}%
\centering{}\includegraphics[width=80mm]{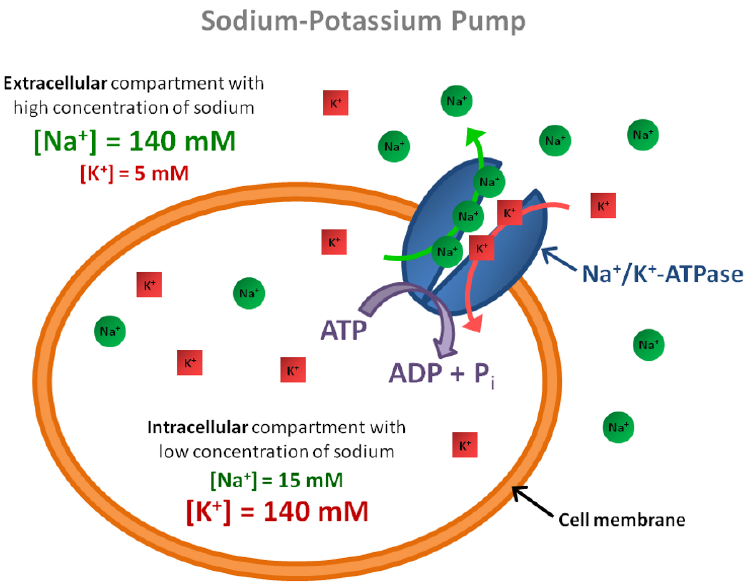}\caption{\label{fig:sodium_pump}Schematics of the sodium-potassium pump (Na$^{+}$/K$^{+}$-ATPase).}
\end{wrapfigure}

The Na$^{+}$/K$^{+}$-ATPase (also called sodium-potassium pump,
or just sodium pump) is present within the membrane of every animal
cell \cite{1994_40_9,1992_24_3}. It is a plasma membrane-associated
protein complex that is expressed in most eukaryotic cells and can
be considered either as an enzyme (ATPase) or as an ion transporter
(the Na$^{+}$-pump). Its main function is to maintain the sodium
and potassium gradients across the membrane, and therefore participates
in the resting potential of the cell, by pumping three intracellular
sodium ions out of the cell and two extracellular potassium ions within
the cell (see Fig. \ref{fig:sodium_pump}).

This ion transport is performed against the electrochemical Na$^{+}$
and K$^{+}$ gradients existing across the cell membrane and therefore
requires energy provided by adenosine triphosphate (ATP) hydrolysis.
This high electrochemical gradient is essential to protect the cell
from bursting as a result of osmotic swelling and also creates a potential
that is used for transmitting nerve impulses and for pumping ions
(such as protons, calcium, chloride, and phosphate) and metabolites
and nutrients across the cell membrane (such as glucose and amino
acids) or neurotransmitters (such as glutamate in astrocyte cells).
ATPase activity and ion transport are intimately linked and are two
aspects of the same function. Regulation of Na$^{+}$/K$^{+}$-ATPase
therefore plays a key role in the etiology of some pathological processes.
For example, when the demand for ATP exceeds its production, the ATP
supply for the Na$^{+}$/K$^{+}$-ATPase will be insufficient to maintain
the low intracellular sodium concentration and thus an increase of
intracellular sodium concentration can be observed.

\chapter{Sodium Nuclear Magnetic Resonance}

\section{Short History of Sodium NMR }

\begin{wrapfigure}{o}[30mm]{9cm}%
\begin{centering}
%\vspace{-6mm}
\par\end{centering}

\begin{centering}
\includegraphics[width=8cm]{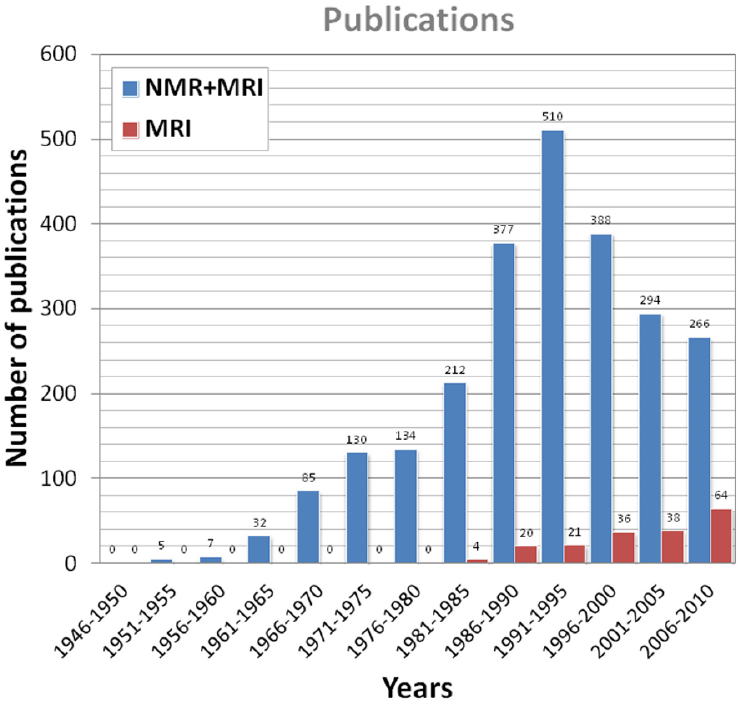}
\par\end{centering}

\caption{\label{fig:Publi_sodium_nmr}Publication statistics on sodium NMR
and MRI. Data was obtained from a research in Google Scholar with
the research key words ``sodium'' or ``23Na'' along with ``magnetic
resonance'' or ``magnetic resonance imaging'' or ``MRI'' or ``NMR''
in the publication title.}
\end{wrapfigure}

The bar chart in Fig. \ref{fig:Publi_sodium_nmr} presents the distribution
of publications on sodium NMR and MRI over the years since the discovery
of magnetic resonance in the 1940s \cite{1946_70_7,1946_69_1}. Data
was obtained from a research in Google Scholar with the research key
words ``sodium'' or ``23Na'' along with ``magnetic resonance''
or ``magnetic resonance imaging'' or ``MRI'' or ``NMR'' in the
publication title.

It is interesting to notice the peak of publications in the period
1991-1995 for the combination of sodium NMR (spectroscopy) and MRI,
which then decreases while the publications on sodium MRI alone continues
to increases since the early 1980s. A shift of interest appears from
sodium NMR spectroscopy to MRI, which is probably due to the recent
availability of scanners with high fields and therefore the possibility
of performing sodium MRI \emph{in vivo} in reasonable times, opening
to new exciting research perspectives.

Briefly, biological tissues were already investigated with sodium
NMR spectroscopy in the early 1970s \cite{1973_204,1973_204_297}
and with sodium MRI in the early 1980s \cite{1985_156_1,1984_40_2,1985_3_4},
first on animals \emph{in vivo} and then on human brain \cite{1985_9_1}
and human heart and abdomen \cite{1988_7_1}. Sodium MRI was therefore
applied for brain tumor and ischemia detection in the late 1980s \cite{1988_30_5}.
The years 1990s will see an increase of interest in sodium MRI, due
to the increase of the magnetic fields in scanners, improvements of
electronics and RF coils, and new rapid sequences that allow to acquire
sodium images in a few minutes with a reasonable resolution of a few
millimeters \cite{1997_37_5}, or new contrasts such as triple quantum
filtered images \cite{1991_95_2,1999_141_2,1999_42_6}. This trend
continued in the 2000s until today.

\section{Sodium NMR Properties}

The sodium nucleus has a spin $I=\nicefrac{\mathsf{3}}{\mathsf{2}}$
and possesses a quadrupolar moment Q which interacts with the electric
field gradients (EFG) generated by the electronic distribution around
the nucleus. In general, sodium in solids experience quadrupolar interactions
while in liquids, the static quadrupolar interaction is averaged to
zero. In the intermediate regimes, e.g. in biological tissues, the
quadrupolar interaction results in a biexponential relaxation behavior.
The dominance of quadrupolar relaxation mechanism for NMR signals
can allow a sensitive characterization of the molecular environment
of the sodium ions. A short T2 component T2$_{\mathsf{short}}$ =
0.5-5 ms generally contributes to 60\% of the signal in living tissue
and a long component T2$_{\mathsf{long}}$ = 15-30 ms corresponds
to 40\% of the signal. In order to detect both T2 components, imaging
techniques with ultrashort echo time (UTE) of less than 0.5 ms are
required. But it is difficult to obtain reliable information about
the biexponential relaxation from these conventional simple single-pulse
/ single quantum (SQ) NMR methods. It has been shown that in the presence
of biexponential relaxation, multiple quantum coherences (MQC) can
be created and detected with MQF sequences \cite{1986_85_11}, and
thus give information about the interaction of sodium ions with their
environment. Ranges of sodium concentrations and relaxation times
in some human tissues \emph{in vivo} are given in Table \ref{tab:sodium_relaxation_times}

\begin{table}[h]
\caption{\label{tab:sodium_relaxation_times}Range of sodium concentrations
and relaxation times in some human tissues \emph{in vivo}.~~~~~~~~~~~~~~~~~~~~~~~~~~~~~~~~~~~~~~}

\centering{}%
\begin{tabular*}{1\columnwidth}{@{\extracolsep{\fill}}llllll}
\hline 
\textbf{\footnotesize Tissue} &  & \textbf{\footnotesize {[}Na$^{+}${]} (mM)} & \textbf{\footnotesize T1 (ms)} & \textbf{\footnotesize T2$_{\mathsf{short}}$ (ms)} & \textbf{\footnotesize T2$_{\mathsf{long}}$ (ms)}\tabularnewline
\hline 
\textbf{\footnotesize Brain} & \textbf{\footnotesize WM} & {\footnotesize 20-60} & {\footnotesize 15-35} & {\footnotesize 0.8-3} & {\footnotesize 15-30}\tabularnewline
 & \textbf{\footnotesize GM} & {\footnotesize 30-70} & {\footnotesize 15-35} & {\footnotesize 0.8-3} & {\footnotesize 15-30}\tabularnewline
 & \textbf{\footnotesize CSF} & {\footnotesize 140-150} & {\footnotesize 50-55} & {\footnotesize -} & {\footnotesize 55-65}\tabularnewline
\textbf{\footnotesize Cartilage} &  & {\footnotesize 250-350} & {\footnotesize 15-25} & {\footnotesize 0.5-2.5} & {\footnotesize 10-30}\tabularnewline
\textbf{\footnotesize Blood} &  & {\footnotesize 140-150} & {\footnotesize 20-40} & {\footnotesize 2-3} & {\footnotesize 12-20}\tabularnewline
\textbf{\footnotesize Muscle} &  & {\footnotesize 15-30} & {\footnotesize 12-25} & {\footnotesize 1.5-2.5} & {\footnotesize 15-30}\tabularnewline
\hline 
\end{tabular*}
\end{table}

\section{Tensors and Hamiltonians }

\subsection*{Irreducible Spherical Tensor Operator (ISTO)}

A convenient way to describe the various processes occurring during
the pulse sequence is by the irreducible spherical tensor operator
(ISTO) basis \cite{1986_67_3_1,1986_67_3_2,2003_19A_2I}, $T_{nm}$,
where $n$ is the rank and $m$ is the order (or coherence) of the
tensor with $\left|m\right|<n$. These tensors are composed of combinations
of the more familiar spin operators $I_{z}$ and $I_{\pm}=I_{x}\pm iI_{y}$.
The ISTO used to describe the density operator and the Hamiltonians
for a spin $\nicefrac{\mathsf{3}}{\mathsf{2}}$ are given in Table
\ref{tab:ISTO}. In order to follow the sequence of events at high
magnetic fields, one should bear in mind two simple rules: 
\begin{enumerate}
\item A non-selective radiofrequency (RF) pulse can change the coherence
$m$ within the limits of $\left|m\right|<n$.
\item Changes in the rank $n$ can occur by relaxation and modulation by
quadrupolar, dipolar or J-coupling, keeping the same value of $m$. 
\end{enumerate}
In the following sections, we consider that the spin system evolves
only under the influence of Zeeman, quadrupolar and RF Hamiltonians
\cite{2003_19A_2I,2005_46_63}.

\begin{table}[h]
\caption{\label{tab:ISTO}Irreducible spherical tensor operators (ISTO) for
spin $I=\nicefrac{\mathsf{3}}{\mathsf{2}}$.~~~~~~~~~~~~~~~~~~~~~~~~~~~~~~~~~~~~~~~~~~~~~~~~~~~~~~~~~~~~~~~~~~~~~~~~~~~~~~~~~~~~}

\begin{centering}
\begin{tabular*}{1\columnwidth}{@{\extracolsep{\fill}}lll}
\hline 
\noalign{\vskip\doublerulesep}
\textbf{\small ISTO} & \textbf{\small Value} & \textbf{\small Definition}\tabularnewline[\doublerulesep]
\hline 
\noalign{\vskip\doublerulesep}
\noalign{\vskip\doublerulesep}
{\small $T_{00}$} & {\small $1$} & {\small Identity}\tabularnewline[\doublerulesep]
\noalign{\vskip\doublerulesep}
\noalign{\vskip\doublerulesep}
{\small $T_{10}$} & {\small $I_{z}$} & {\small Longitudinal magnetization}\tabularnewline[\doublerulesep]
\noalign{\vskip\doublerulesep}
\noalign{\vskip\doublerulesep}
{\small $T_{1\pm1}$} & {\small $\mp\frac{1}{\sqrt{2}}I_{\pm}$} & {\small Rank 1 SQC}\tabularnewline[\doublerulesep]
\noalign{\vskip\doublerulesep}
\noalign{\vskip\doublerulesep}
{\small $T_{20}$} & {\small $\frac{1}{\sqrt{6}}\left(3I_{z}^{2}-I\left(I+1\right)\right)$} & {\small Quadrupolar magnetization}\tabularnewline[\doublerulesep]
\noalign{\vskip\doublerulesep}
\noalign{\vskip\doublerulesep}
{\small $T_{2\pm1}$} & {\small $\mp\frac{1}{2}\left[I_{z},I_{\pm}\right]_{+}$} & {\small Rank 2 SQC}\tabularnewline[\doublerulesep]
\noalign{\vskip\doublerulesep}
\noalign{\vskip\doublerulesep}
{\small $T_{2\pm2}$} & {\small $\frac{1}{2}I_{\pm}^{2}$} & {\small Rank 2 DQC}\tabularnewline[\doublerulesep]
\noalign{\vskip\doublerulesep}
\noalign{\vskip\doublerulesep}
{\small $T_{30}$} & {\small $\frac{1}{\sqrt{10}}\left(5I_{z}^{3}-\left(3I\left(I+1\right)-1\right)I_{z}\right)$} & {\small Octopolar magnetization}\tabularnewline[\doublerulesep]
\noalign{\vskip\doublerulesep}
\noalign{\vskip\doublerulesep}
{\small $T_{3\pm1}$} & {\small $\mp\frac{1}{4}\sqrt{\frac{3}{10}}\left[5I_{z}^{3}-I\left(I+1\right)-\frac{1}{2},I_{\pm}\right]_{+}$} & {\small Rank 3 SQC}\tabularnewline[\doublerulesep]
\noalign{\vskip\doublerulesep}
\noalign{\vskip\doublerulesep}
{\small $T_{3\pm2}$} & {\small $\frac{\sqrt{3}}{4}\left[I_{z},I_{\pm}^{2}\right]_{+}$} & {\small Rank 3 DQC}\tabularnewline[\doublerulesep]
\noalign{\vskip\doublerulesep}
\noalign{\vskip\doublerulesep}
{\small $T_{3\pm3}$} & {\small $\mp\frac{1}{2\sqrt{2}}I_{\pm}^{3}$} & {\small Rank 3 TQC}\tabularnewline[\doublerulesep]
\hline 
\noalign{\vskip\doublerulesep}
\end{tabular*}
\par\end{centering}

{\scriptsize ~}{\scriptsize \par}

{\scriptsize SQC, DQC and TQC are respectively the single, double
and triple quantum coherences. }{\scriptsize \par}

{\scriptsize The anticommutator for the operators A and B is defined
as $\left[A,B\right]_{+}=AB+BA$.}
\end{table}

\subsection*{Zeeman Hamiltonian $(H_{z})$}

The Zeeman Hamiltonian is given by 
\begin{equation}
H_{z}=\omega_{0}T_{10},\label{eq:Hz}
\end{equation}
with $\omega_{0}=\gamma_{Na}B_{0}$ the Larmor angular frequency,
$\gamma_{Na}=70.808493\times10^{6}$ $rad\cdot T^{-1}\cdot s^{-1}$
the sodium gyromagnetic ratio (26\% of the proton $\gamma_{p}$) and
$B_{0}$ the magnetic field. 

~\\
All the following Hamiltonians are expressed in the Larmor frequency
rotating frame, in which this $H_{z}$ vanishes.

\subsection*{Quadrupolar Hamiltonian $(H_{Q})$}

\subsubsection*{Electric field gradient (EFG)}

The nuclear quadrupole interaction is determined by the orientation,
magnitude and temporal duration of EFG generated by the electronic
configuration of the molecular environment surrounding the nucleus.
If $V(x,y,z)$ is the electrostatic potential produced by electrons
at the point (x,y,z), EFG can be described by the tensor $V_{\alpha\beta}$
with the components:
\begin{equation}
V_{\alpha\beta}=\frac{\partial^{2}V}{\partial\alpha\partial\beta},\qquad(with\,\alpha,\beta=x,y,z).\label{eq:EFG_V}
\end{equation}

If we choose the principal axes of the symmetric tensor $V_{\alpha\beta}$
as coordinates axes, the cross-terms vanish, i.e. $V_{xy}=V_{xz}=V_{yz}=0$.
It is common to label the remaining components such that $\left|V_{zz}\right|\geq\left|V_{yy}\right|\geq\left|V_{xx}\right|$,
and to define $V_{zz}=eq$ (e is the unit electric charge) and the
anisotropy parameter:
\begin{equation}
\eta=\frac{V_{xx}-V_{yy}}{V_{zz}},\qquad0\leq\eta\leq1.\label{eq:eta}
\end{equation}

\subsubsection*{Quadrupolar Hamiltonian }

The quadrupolar interaction of the $\mathsf{^{23}Na}$ spin with the
EFG is represented by the quadrupolar Hamiltonian $H_{Q}$. In the
principal axis system (PAS) of $V_{\alpha\beta}$, it reads: 
\begin{equation}
H_{Q}=\omega_{Q}\sum_{m=-2}^{2}\left(-1\right)^{m}F_{2,-m}T_{2m},\label{eq:HQ}
\end{equation}
with the PAS spatial tensor components: 
\begin{equation}
\begin{array}{lcl}
F_{2,0} & = & \sqrt{\frac{3}{2}},\\
F_{2,\pm1} & = & 0,\\
F_{2,\pm2} & = & \frac{\eta}{2},
\end{array}\label{eq:F_PAS}
\end{equation}
and the quadrupolar angular frequency:
\begin{equation}
\omega_{Q}=\frac{e^{2}qQ}{6\hbar}.\label{eq:omegaQ}
\end{equation}

\subsection*{Radiofrequency Hamiltonian $(H_{RF})$}

If the RF pulse is applied on-resonance along the $x$-axis with field
strength $\omega_{RF}\left(t\right)=\gamma_{Na}B_{1}\left(t\right)$,
with $B_{1}\left(t\right)$ the RF magnetic field, the corresponding
RF Hamiltonian is time-dependent and reads:
\begin{equation}
H_{RF}=\omega_{RF}I_{x}=\omega_{RF}\frac{1}{\sqrt{2}}\left(T_{1-1}-T_{1+1}\right).\label{eq:HRF}
\end{equation}

\subsection*{Total Hamiltonian $(H)$}

In the Larmor rotating frame, the total Hamiltonian is then:
\begin{equation}
H=H_{RF}+H_{Q}.\label{eq:H_total}
\end{equation}

\section{Quadrupolar Relaxation and NMR Spectra}

\begin{figure}[h]
\begin{centering}
\includegraphics[width=1\columnwidth]{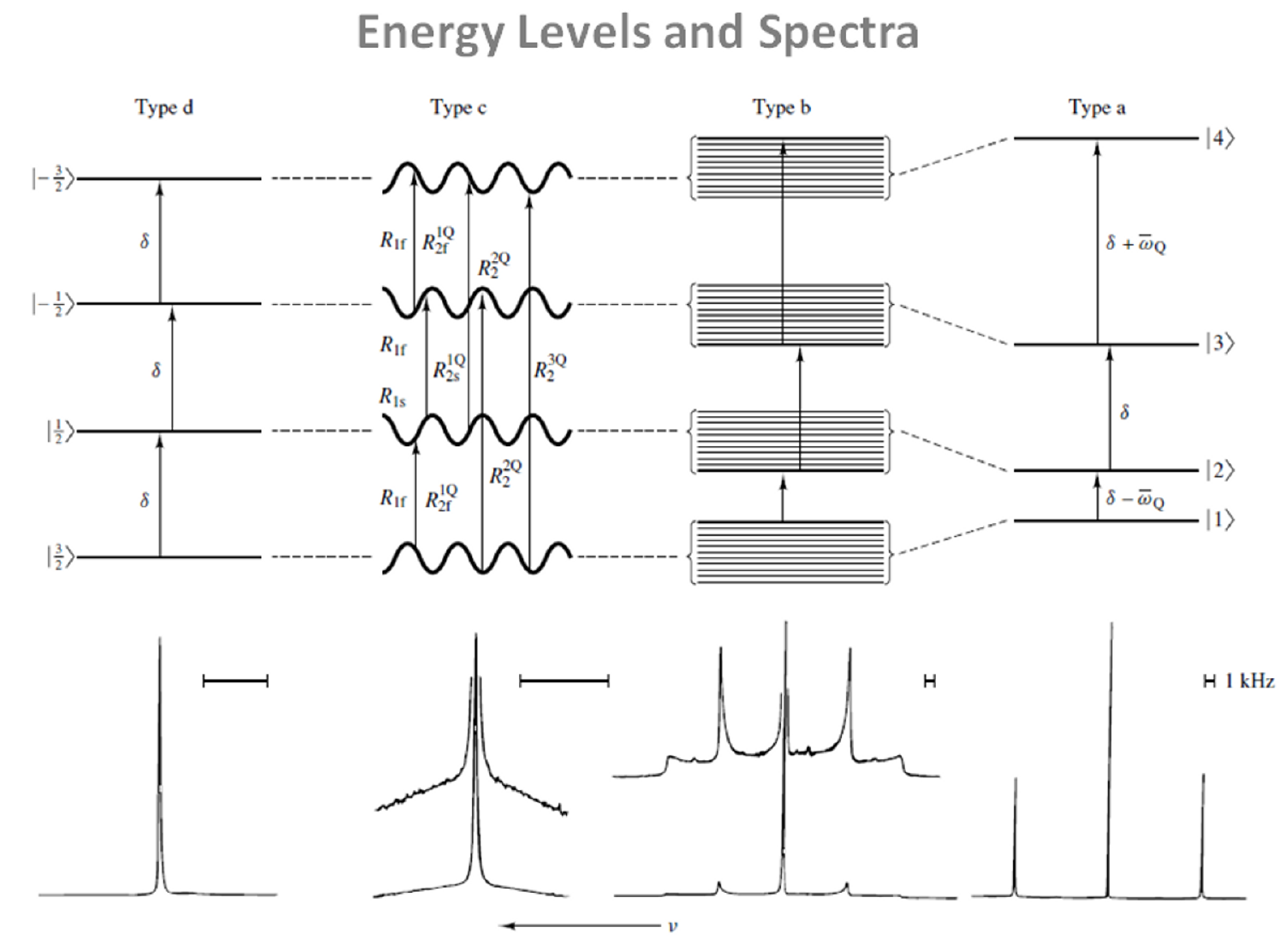}
\par\end{centering}

\caption{\label{fig:NMR_spectra}Typical energy levels and corresponding NMR
spectra of sodium nuclei in different environments. Four types of
motionally narrowed SQ spectra are possible (a, b, c, and d). The
\textbf{type-a} (crystal-like) and \textbf{type-b} (powder-like) spectra
are of Na+ in aqueous suspensions of oriented and unoriented dodecyl
sulfate micelles, respectively. The \textbf{type-c} (homogeneous,
biexponential, super-lorentzian-like) spectrum is that of Na+ in an
aqueous solution that has a high concentration of micelle-solubilized
gramicidin channels. The \textbf{type-d} (liquid-like) spectrum is
that of NaCl in H2O. From Rooney W. et al. NMR in Biomedicine 4, 209-226,
1991. Reproduced by permission of Wiley-Blackwell.}
\end{figure}

Depending on the motional regime of the system, the energy levels
of the system and the relaxation rates can be modified, giving one
or many NMR peaks in the one-pulse single quantum (SQ) $\mathsf{^{23}Na}$
spectra, as shown on Fig. \ref{fig:NMR_spectra} \cite{1991_4_5_209}.
The following three motional regimes and four different types of spectra
are possible, depending on the nature of the spin dynamics and the
nature of the molecular environment \cite{1997_31,1991_4_5_227,1991_4_5_209,2006_19_7}.
In the following section, we consider that the Na$^{+}$ ions are
present in a single compartment with a single rotational correlation
time $\tau_{c}$. We also consider that the system is in the Redfield
regime, where the relaxation of spin \textrm{$\nicefrac{\mathsf{3}}{\mathsf{2}}$}
is described by second-order perturbation theory, and where the spin
dynamics can be solved analytically \cite{2003_19A_2I}.

\subsection*{Isotropic Motion With Motional Narrowing $(\omega_{0}\tau_{c}\ll1)$}

In a system of rapid motion such as a fluid, all orientations of the
EFG are equally probable. In this isotropic system, the quadrupolar
interaction is averaged to zero on the time scale of $\nicefrac{2\pi}{\omega_{0}}$
and the four energy levels of the spin $\nicefrac{3}{2}$ are equally
distant by the angular frequency $\omega_{0}$. The $^{\mathsf{23}}$Na
spectrum is then composed of a single resonance line at $\omega_{0}$,
as shown in the type-d spectrum in Fig. \ref{fig:NMR_spectra}. Both
transverse and longitudinal relaxations are simple exponential decays.
The liquid-like type-d spectrum was obtained from NaCl in aqueous
solution.

\subsection*{Isotropic Motion Without Motional Narrowing $(\omega_{0}\tau_{c}\sim1)$}

In tissues where the macromolecular motion associated with the nucleus
is isotropic but of the same order of the Larmor frequency, the quadrupolar
interaction dominates the relaxation. The satellite and central transitions
have different transverse relaxation rates $R_{1}^{\left(\pm1\right)}$
and $R_{2}^{\left(\pm1\right)}$ respectively, and are dynamically
shifted from the Larmor frequency with shifts $K_{1}$ and $K_{2}-K_{1}$,
respectively. However, these shifts are much smaller than the linewidths
and very difficult to detect. This situation can correspond to a biological-like
type-c spectrum in Fig. \ref{fig:NMR_spectra}, which was obtained
from solubilized micelles in an aqueous solution. It is a ``homogeneous''
(``biexponential'', ``super-Lorentzian'') spectrum, in which the
satellite peaks have essentially coalesced into a single, broad, homogeneous
peak. The relaxation rates are:
\begin{equation}
\begin{array}{lcl}
R_{1}^{\left(\pm1\right)} & = & J_{o}+J_{1}\pm iK_{1},\\
R_{2}^{\left(\pm1\right)} & = & J_{1}+J_{2}\mp i\left(K_{1}-K_{2}\right)
\end{array}\label{eq:R1_R2_1}
\end{equation}
with the spectral densities (with $m=0,1,2$):
\begin{equation}
\begin{array}{lcl}
J_{m}\left(m\omega\right) & = & \frac{\left(2\pi\right)^{2}}{20}\left(\frac{\chi^{2}\tau_{c}}{1+\left(m\omega\right)^{2}\tau_{c}^{2}}\right),\\
K_{m}\left(m\omega\right) & = & \omega\tau_{c}J_{m}\left(m\omega\right).
\end{array}\label{eq:Jm_Km}
\end{equation}
The spectral densities $J_{m}$ and $K_{m}$ are the real and imaginary
parts of the FFT of the EFG correlation function and $\chi$ is the
root mean square coupling constant. Even in the absence of any quadrupolar
splitting, biexponential relaxation can create MQC which can be detected
by triple quantum filtering (TQF).

\subsection*{Anisotropic Motion $(\omega_{0}\tau_{c}>1)$}

The sodium nuclei experiment non-averaged EFG and therefore a residual
quadrupolar interaction (RQI). The SQ relaxation rates are:
\begin{equation}
\begin{array}{lcl}
R_{1}^{\left(1\right)} & = & J_{0}+J_{1}+J_{2}-\sqrt{J_{2}^{2}-\omega_{Q}^{2}},\\
R_{2}^{\left(1\right)} & = & J_{1}+J_{2},\\
R_{3}^{\left(1\right)} & = & J_{0}+J_{1}+J_{2}+\sqrt{J_{2}^{2}-\omega_{Q}^{2}}.
\end{array}\label{eq:R1_R2_2}
\end{equation}
Depending on the magnitude of $J_{2}$, three situations can occur:
\begin{enumerate}
\item If $\omega_{Q}<J_{2}$, the relaxation rates are real and there is
no line splitting despite the presence of RQI. The SQ spectrum is
a sum of three Lorentzians. The RQI influences the linewidths and
amplitudes of the components, there is a very small splitting in the
spectrum, but it is masked by the larger SQ linewidths. This case
can be still described by a type-c spectrum in Fig. \ref{fig:NMR_spectra}.
In this case, the Double Quantum filter (DQF) signal due to even rank
coherence $T_{2\pm2}$ can be used to detect and measure $\omega_{Q}$.
\item If $\omega_{Q}>J_{2}$, the relaxation eigenvalues and corresponding
to outer transitions are complex and the satellite transitions are
shifted by $\pm\sqrt{\omega_{Q}^{2}-J_{2}^{2}}$ from the central
line. This case corresponds to a powder-like type-b spectrum in Fig.
\ref{fig:NMR_spectra}, which was acquired from unoriented micelles
in aqueous solution (unoriented liquid crystal). The DQF signal from
$T_{2\pm2}$ can also be used to measure $\omega_{Q}$.
\item If $\omega_{Q}\gg J_{2}$, the energy levels are all shifted by the
RQI resulting in three distinct peaks in the spectrum. The central
transition and the two satellite transitions are separated by $\omega_{Q}$.
This frequency separation between the lines provides therefore indirect
information about RQI and the magnitude of ordering in the system.
This case can be described by a crystal-like type-a spectrum in Fig.
\ref{fig:NMR_spectra}, which was obtained from oriented micelles
in aqueous solution (oriented liquid crystal). Relative areas of the
peaks are 3:4:3.
\end{enumerate}

\section{Spectra Models}

In tissue, Na$^{+}$ aquo cations exist in compartmentalized spaces
(intra- and extracellular compartments for example) and encounter
an abundance of charged macromolecules. The nature of tissue is such
that most of the sodium spectra are likely to be of type c or type
b, or intermediate or superpositions of these. Several models were
therefore developed in order to understand the SQ sodium spectra in
biological tissues and are described in more details in Ref. \cite{1991_4_5_209,1991_4_5_227,2007_EMR}
and references herein.

\subsection*{Debye Model (DM)}

The DM is the simplest model for the quadrupolar relaxation mechanism
and is based on the assumption that the random fluctuation of the
EFG tensor strength and orientation that can be described with a single
correlation time, $\tau_{c}$. This is the model that was used to
describe the spectra in the previous section (quadrupolar relaxation),
assuming that the sodium ions are all in a single environmental compartment
and that no exchange between compartments occurs. However, this model
is too simple to describe sodium spectra in even simple biological
tissues such as bovine serum albumin in saline solution or suspensions
of yeast cells.

\subsection*{Discrete Exchange Model (DEM)}

The DEM is a more complex model based on equilibrium chemical exchange
between distinct sites that have different quadrupolar properties.
These individual sites can be each characterized by a Debye model
(single $\tau_{c}$). For example, the James-Noggle exchange occurs
between two different type-d sites and can yield only a resultant
type-d spectrum. Bull exchange happens between a type-d site and a
type-c site and can yield either type of resultant spectrum, depending
on the lability of the exchange on the timescale determined by the
properties of the two sites. Chan exchange occurs between a type-d
site and a type-a site and can produce all kinds of resultant spectra,
depending on the lability of the exchange or the random orientation
distribution of the type-a sites. However, a two-site DEM model has
at least six (or even eight, if temperature is included) independent
parameters to describe the resultants spectra, and a three-site exchange
model requires at least 13 independent parameters. Analyzing the tissue
situation with a realistic DEM is therefore very difficult.

\subsection*{Berendsen-Edzes Model (BEM)}

The BEM seems is a more realistic model which focuses on the EFG tensor
projection fluctuations caused by motions of the inner hydration shell
of the ion Na$^{+}$(aq), that are very powerful and very rapid. Slower
modulations also occur, usually of lower amplitudes, and are generated
as the aquo ion diffuses and encounters macromolecules. The BEM mechanism
is based on the concept of a sample domain, as experienced by a diffusing
aquo ions, which must be at least as large as the average volume sampled
by the diffusional excursions during the lifetime of an NMR coherence.
The BEM can therefore be useful for describing spectra by modeling
domains with the same (type-a) or random (type-b) orientations. A
type-c spectrum can also be described using BEM, for Na$^{+}$ in
the blood plasma for example, where rapid modulations of the inner
hydration sphere domain happen altogether with isotropic slow fluctuations
of the EFG by diffusion. But this model is still very difficult to
interpret in complex biological tissues.

\subsection*{Other Models }

Some other more complex models such as continuous diffusion models,
models with distributions of $\tau_{c}$ values, or Debye models with
a distribution of $\omega_{Q}$ values were also developed in order
to describe more accurately sodium spectra in biological tissues.

\subsection*{Rooney-Springer Protocol}

In order to really have a accurate understanding of the sodium NMR
data acquisitions in biological tissues, Rooney and Springer \cite{1991_4_5_209}
proposed a protocol in five steps. Although their complete realization
is very difficult, if not impossible, it is still important to consider
them and to try to complete at least some of them in order to have
a good approximate understanding of the results we obtain from sodium
NMR spectroscopy or MRI \emph{in vivo}: 
\begin{enumerate}
\item \vspace{-3mm}Discriminate the SQ resonances from as many physiological
compartments as possible. 
\item \vspace{-3mm}Discriminate the spectral type (a, b, c or d) of each
resonance. 
\item \vspace{-3mm}Determine the numbers of magnetically distinct populations
in slow exchange for each resonance (the type of which has been determined
above). 
\item \vspace{-3mm}Measure as many relaxation times values as possible
for each population. 
\item \vspace{-3mm}Interpret the measured relaxation times values.
\end{enumerate}

\section{Multiple Quantum Filters (MQF)}

In order to detect MQC from different types of spectrum and therefore
obtain some information about the biological environment of the sodium
nuclei, MQF can be implemented before the NMR acquisitions \cite{1986_85_11,2001_14_2,1990_88_2,1980_14}.
MQF are described in Fig. 4. For all MQF experiments, the RF pulses
were considered as hard pulses ($\omega_{RF}\gg\omega_{Q}$).

A first RF pulse of angle $\theta_{1}$ and phase $\phi_{1}$ is applied
to the system to flip the longitudinal magnetization $T_{10}$ into
the transverse plane, giving a transverse magnetization $T_{1\pm1}$.
During the preparation time $\tau$, the transverse magnetization
$T_{1\pm1}$ evolves in the presence of relaxation and/or RQI, creating
single quantum coherences (SQC) with different rank $T_{1\pm1}$,
$T_{2\pm1}$, $T_{3\pm1}$. \begin{wrapfigure}{o}[30mm]{90mm}%
\centering{}\includegraphics[width=90mm]{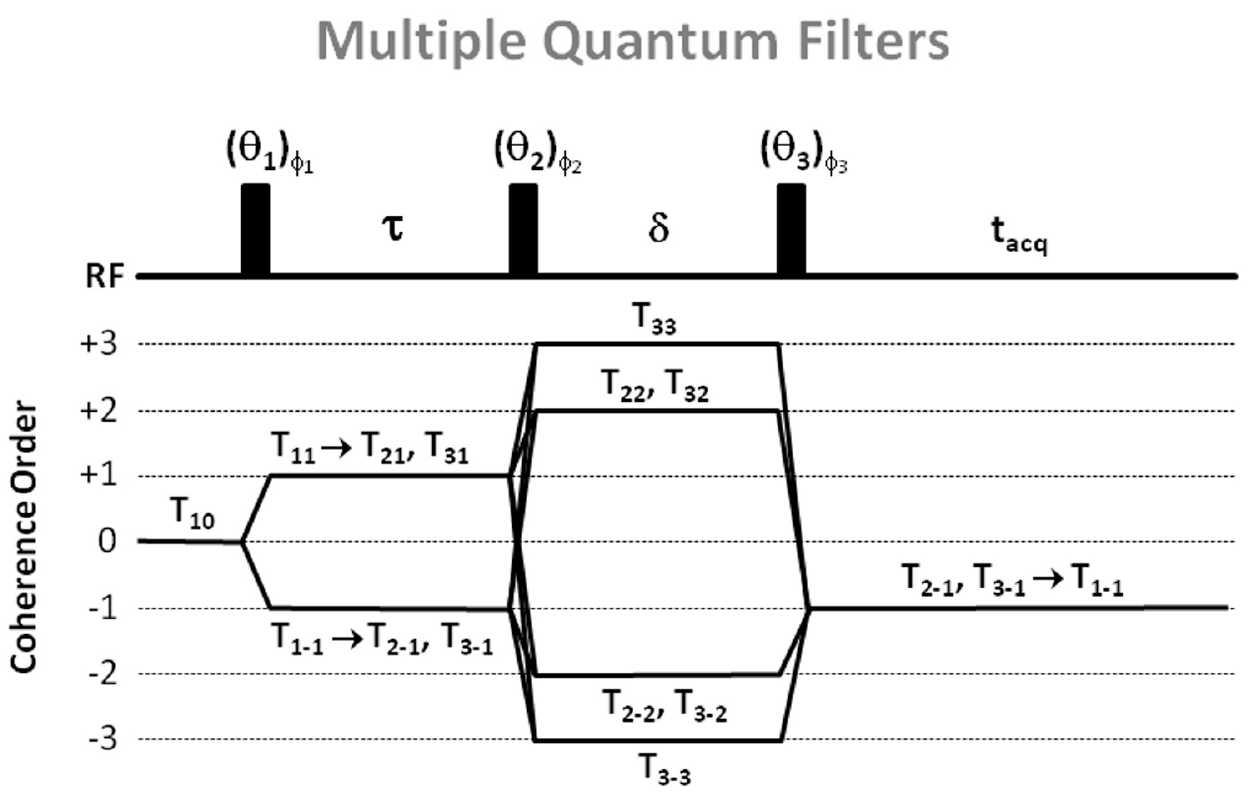}\caption{\label{fig:MQF}Multiple quantum filters (MQF) and coherence order
transfers. For triple quantum filter (TQF) with six-phase cycling,
$\theta_{1}=\theta_{2}=\theta_{3}=90^{\circ}$, $\phi_{1}=30^{\circ},\,90^{\circ},\,150^{\circ},\,210^{\circ},\,270^{\circ},\,300^{\circ}$,
$\phi_{2}=\phi_{1}+90^{\circ}$, $\phi_{3}=0^{\circ}$. For double
quantum filter (DQF) with four-phase cycling, $\theta_{1}=\theta_{2}=\theta_{3}=90^{\circ}$
, $\phi_{1}=0^{\circ},\,90^{\circ},\,180^{\circ},\,270^{\circ}$,
$\phi_{2}=\phi_{1}$, $\phi_{3}=0^{\circ}$. For DQF with magic angle
(DQF-MA), $\theta_{1}=90^{\circ}$, $\theta_{2}=\theta_{3}=54.7^{\circ}$
, and the phase cycling is the same as for DQF. For all filters, the
receiver phase alternates between $0^{\circ}$ and $180^{\circ}$
.}
\end{wrapfigure}%
In isotropic liquids only odd rank tensors $T_{3\pm1}$ can be formed
during $\tau$ under the influence of quadrupolar relaxation. In anisotropic
media, where the RQI does not average to zero, rank tensors $T_{2\pm1}$
can be also formed. 

A second pulse $\theta_{2}$ and phase $\phi_{2}$ converts these
SQC into multiple quantum coherences (MQC) $T_{1\pm1}$, $T_{2\pm2}$,
$T_{3\pm2}$, $T_{3\pm3}$. During the evolution time $\delta$, the
MQCs evolve with their characteristic relaxation times. Generally,
the evolution time $\delta$ is kept very short to avoid their decay. 

Since the MQCs cannot be detected directly, the final RF pulse $\theta_{3}$
and phase $\phi_{3}$ converts the MQCs into SQCs $T_{1\pm1}$, $T_{2\pm1}$,
$T_{3\pm1}$, which then evolve under relaxation and RQI to the detectable
SQC $T_{1\pm1}$ during the acquisition time $t_{acq}$. It can be
notice that in general, a $180^{\circ}$ RF pulse can be applied in
the middle of the preparation time $\tau$ in order to refocus the
chemical shifts and magnetic field inhomogeneities. However, as MQF
are very sensitive to RF imperfections, it has been shown that filters
without this refocusing pulse generate a better signal and are more
robust to RF inhomogeneities for MQF MRI \cite{1994_104_2}. The formation
of DQC and TQC is detected separately by choosing the appropriate
RF pulse flip angles and phase cycling. The phase cycling and choice
of RF pulses are given in the caption of Fig. \ref{fig:MQF}. DQF
can detect the contribution of two tensors, $T_{2\pm1}$, $T_{3\pm1}$,
due to the slow motion regime or/and RQI in anisotropic media. if
$\theta_{1}=\theta_{2}=\theta_{3}=90^{\circ}$, contributions from
both $T_{2\pm1}$ and $T_{3\pm1}$ can be detected. If $\theta_{1}=90^{\circ}$
and $\theta_{2}=\theta_{3}=54.7^{\circ}$ (magic angle), only the
contribution of $T_{2\pm1}$ is detected, which arises only from the
anisotropic RQI. Only $T_{3\pm1}$ due to the slow motion regime can
be detected by the TQF sequence.

\section{Fluid and Extracellular Sodium Suppression}

In many diseases, a sodium concentration increases is detected, which
can be caused by either an increase of intracellular sodium concentration,
increase of extracellular volume with a constant concentration (140
mM) or increase of vascularization. It is widely believed that the
most sensitive way to study the health of tissues \emph{in vivo} should
be done by isolating the sodium NMR signal from the intracellular
compartment. Intracellular sodium concentration and relaxation properties
should give access to some more useful information on the cells viability
(homeostasis, energetic state and sodium pump function) \cite{1991_3}.
Four different techniques have been suggested to suppress extracellular
sodium and/or the signal from fluids around the tissue of interest.

\subsection*{Shift Reagents (SR) }

This technique is based on the use a $^{\mathsf{23}}$Na chemical
shift reagent based on lanthanide chelates as Tm(DOTP)$^{\mathsf{5-}}$,
Dy(PPP)$_{2}^{\mathsf{7-}}$ or Dy(TTHA)$^{\mathsf{3-}}$ \cite{1984_13,1987_52_4,2001_45_3}.
These compounds do not penetrate the cell membrane and therefore create
a frequency offset for the sodium nuclei in the extracellular space.
But as they don't cross the blood-brain barrier and because of their
moderate toxicity, these SR can't be used in humans.

\subsection*{Diffusion}

Diffusion-based techniques can separate the sodium signal from the
intracellular and extracellular compartments based on the differences
between the motional properties of the ions in these two compartments
\cite{1993_29_4}. However, the fast T2 relaxation of sodium and its
low gyromagnetic ratio require the use of very large magnetic field
gradients, which can cannot to be effectively implemented in clinical
MRI scanners.

\subsection*{Inversion Recovery (IR)}

The IR technique is based on the difference in the T1 relaxation of
the sodium nuclei in different compartments. As the T1 relaxation
time of the extracellular sodium or in fluid can be significantly
longer than the T1 of the intracellular sodium, IR can be used to
eliminate the signal contribution from either environment \cite{2005_54_5_1305,2010_207_1,2000_6_6}.

\subsection*{Multiple Quantum Filters }

The MQF technique is based on the different T2 relaxation properties
of the sodium nuclei in the intracellular and extracellular or fluid
compartments. MQF can be used to differentiate these compartments,
as they use coherence transfer schemes to generate an NMR signal that
is related to the presence of biexponential relaxation. The MQF are
described in the previous section. Depending on the pulse amplitudes
and the phase cycling, these filters can allow the signal coming from
the sodium DQC or TQC to be read by the sequence. However, the signal
intensity following a MQF sequence is a small fraction of the signal
intensity following a single ideal $90^{\circ}$ excitation pulse
(around 10\% has been given as a typical \emph{in vivo} value). 

Several reports estimate that the MQF NMR signal in biological tissues
comes primarily from the intracellular sodium \cite{1993_30_4,1992_98_1,1986_69_3,1987_72_1_159},
but this statement is still controversial and studies on physiological
samples have shown a significant contribution of the extracellular
sodium in MQ-filtered spectra \cite{1991_3,1993_29_1,1989_81_3}.

\subsection*{Applications to Sodium MRI}

IR and the MQF are the two only techniques that can be currently applied
in sodium MRI \emph{in vivo}. Despite the fact that they cannot really
separate completely the different compartments in biological tissues,
due to their complex distribution of spectrum types and relaxation
times (see Fig. \ref{fig:NMR_spectra}), the resulting signals detected
after application of these filters can still contain more signal weighting
from certain compartments (intracellular) compared to the others (extracellular,
fluid) and thus give some new useful information on the health of
the tissue under study.

\section{RF Pulse Sequences }

Due to the short (biexponential) T2 relaxation of sodium \emph{in
vivo}, ultrashort TE (UTE) sequences are recommended to acquire the
images. The first sodium images were acquired with 3D gradient \begin{wrapfigure}{o}[30mm]{60mm}%
\begin{centering}
\includegraphics[width=60mm]{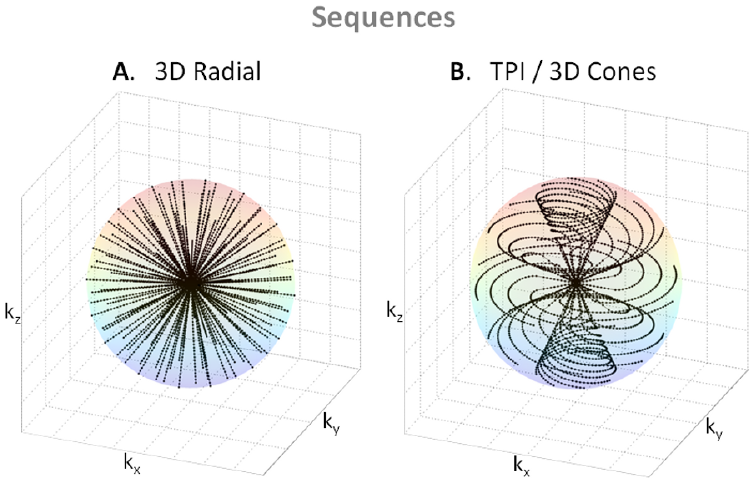}
\par\end{centering}

\caption{\label{fig:kspace}K-space trajectories from UTE sequences. \textbf{A}.
3D radial. \textbf{B}. Twisted projection imaging (TPI) or 3D cones
types of sequences.}
\end{wrapfigure}%
echo sequences \cite{1997_38_4} where the TE was minimized by using
non-selective hard pulse, in order to avoid using a slice refocusing
gradient, and by applying read and phase gradients with the maximum
slew rates and magnitudes. TE of the order of 2-3 ms can be obtained
with this method. The shortest TEs (< 1 ms) can generally be obtained
by acquiring the data from the center of the k-space in a radial or
spiral fashion. The most common type of this kind of sequence is the
3D radial sequence which has been used widely for sodium MRI \cite{2007_57_1}.
This method can then be improved from the signal-to-noise ratio (SNR)
and time efficiency point of view by modifying the density of acquisition
points along the projections and also by twisting the radial projection
in the outer k-space in an optimal manner in order to fill the k-space
more homogeneously and improving the point spread function of the
acquisition method. This is done in sequences such as density adapted
radial sequence \cite{2009_62_6}, twisted projection imaging (TPI)
and its variations \cite{1997_37_5,1997_38_6_1022,2010_63_6_1583},
3D cones \cite{2006_55_3} or FLORET \cite{2011_66_5_1303}. Examples
of k-space trajectories from 3D radial and TPI sequences are shown
in Fig. \ref{fig:kspace}. These latter sequences are now the standard
for sodium MRI. Other UTE sequences have also been developed with
even shorter TEs, but are still under investigation for about their
application to \emph{in vivo} sodium MRI: SPRITE \cite{2006_179_1_64,1996_123_1},
SWIFT \cite{2006_181_2_342}, ZTE \cite{2011_66_2}, PETRA \cite{2012_67_2}.
The data acquired with all these non-Cartesian sequences are then
reconstructed using different methods, such regridding \cite{1985_4_4,2000_19_4,1995_14_3}
or non-uniform fast Fourier transform (NUFFT) algorithms \cite{2007_188_2,2004_43_3_443},
or iterative methods \cite{2007_2007,2007_57_6_1086}.

\section{Sodium Quantification}

Sodium concentration quantification is generally performed by placing
phantoms of known sodium concentration and known relaxation times
within the field-of-view of the images. It is better to use phantoms
with relaxation times that match approximately the relaxation times
of the investigated tissue in order to reduce uncertainties in the
quantification. A linear regression from the phantom signals, corrected
for relaxation, is then used to produce the tissue sodium concentration
(TSC) map. For example, reference phantoms for brain or muscle imaging
are composed of 2-3\% agar gels with sodium concentrations within
the range of TSC usually found in these tissues (10 mM up to 150 mM).
For cartilage, 4-6\% agar gels can be used, with sodium concentration
between 100 to 350 mM.

\section{Limitations and Prospects}

\subsection*{Limitations}

The low sodium concentration in biological tissues compared to the
water concentration, associated with a low NMR sensitivity and very
short transverse relaxation times, make the detection of sodium signal
quite challenging and result in images with low SNR (around 20-40),
low resolution (2-10 mm) and long acquisition times (10-30 min). Moreover,
due the low gyromagnetic ratio of the sodium nuclei (26\% of proton
gyromagnetic ratio), high magnetic field gradients need to be used
for acquiring images with higher resolution, which can therefore be
limited by the hardware capacities of the clinical scanners (around
40 mT/m in general). This gradient limitation can also impede studies
on sodium ions diffusion with MRI due to the very high gradients needed
for detecting even small ion displacements. 

Other limitations include the need of high static magnetic fields
(>3T) for increasing the sodium signal associated with multinuclear
capabilities and the availability of single tuned $^{\mathsf{23}}$Na
coils or dual-tuned $^{\mathsf{1}}$H /$^{\mathsf{23}}$Na RF coils
that allow to acquire co-registered proton and sodium images.

\subsection*{Prospects}

Sodium MRI acquisitions could profit from new recent multichannel
capabilities developed for rapid proton MRI with parallel data acquisition
using array coils and parallel reconstruction algorithms \cite{2010_27_4},
and also new reconstruction schemes such as compressed sensing (CS),
that allow rapid undersampled data acquisitions \cite{2007_58_6_1182}.
A optimized combination of parallel-CS-NUFFT reconstruction associated
with optimized sequences such as TPI at high field could therefore
allow to significantly reduce the acquisition time to a few minutes
and still generate images of good resolution (1-2 mm) and reasonable
SNR, that would put sodium MRI in the realm of practical clinical
imaging techniques.

As assessing the intracellular sodium content can give more useful
metabolic information on the integrity of cells, techniques that increase
the sensitivity of sodium MRI to this concentration. such as IR or
TQF, still need to be improved, in order to increase the SNR, and
thus reduce the resolution and acquisition time, of the images, as
well as reducing their sensitivities to magnetic field and RF inhomogeneities
\cite{2010_23_10}. A recent new method based on optimal control theory
\cite{2010_494_4-6,2009_131_17} could also prove to be useful for
this purpose, but it is still under investigation and has not been
applied to \emph{in vivo} sodium MRI yet.

\chapter{Biomedical Applications}

In this part we will give an overview of the possible biomedical applications
of sodium MRI for assessing diseases and therapies non-invasively
and quantitatively \emph{in vivo}. Of course, this list of applications
is not exhaustive, but we expect it shows a fair spectrum of the research
that has been performed over the last 30 years on this subject.

\section{Brain}

\begin{figure}[h]
\begin{centering}
\includegraphics[width=1\columnwidth]{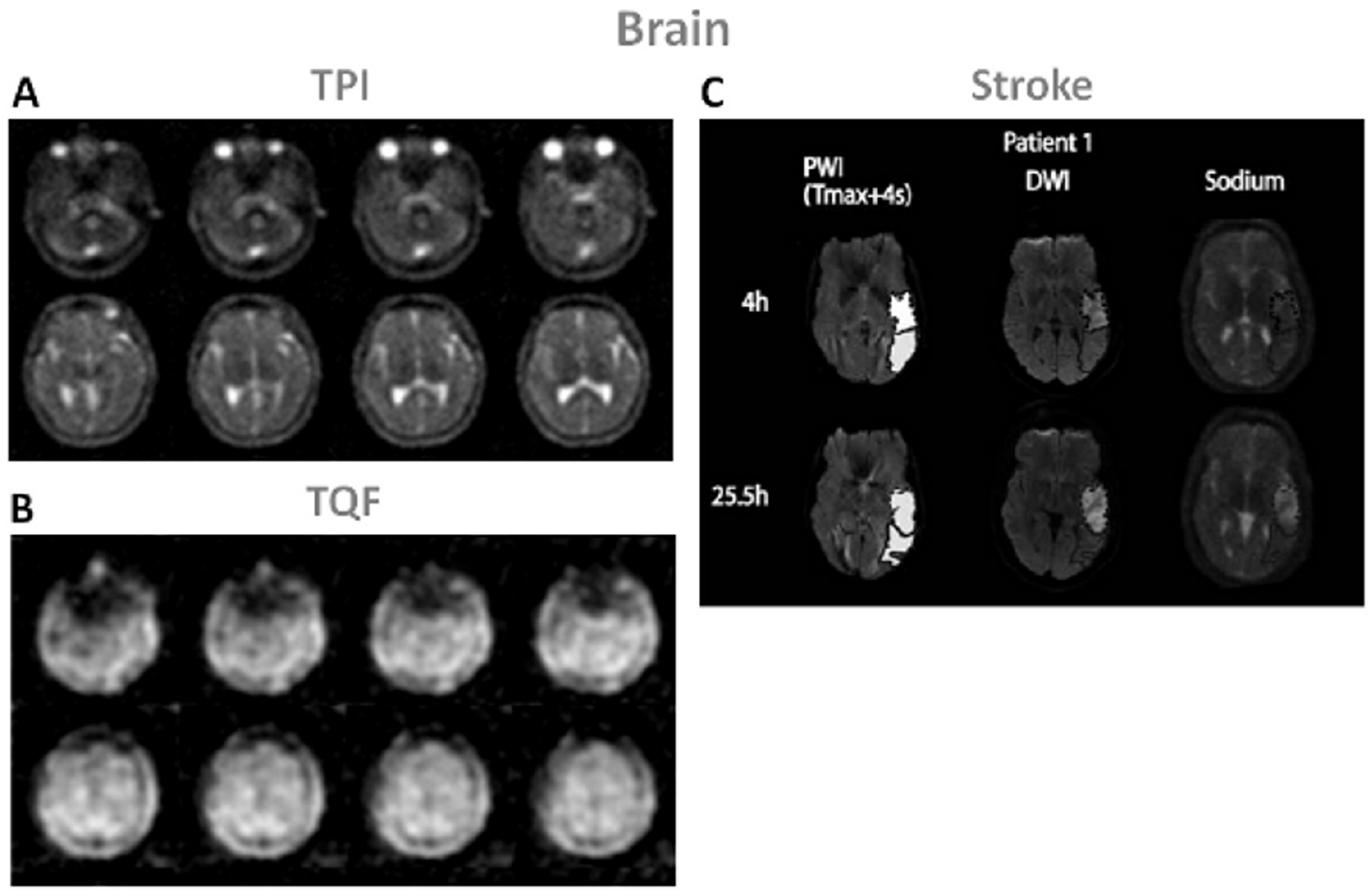}
\par\end{centering}

\caption{\label{fig:brain_1}Examples of brain images. \textbf{A}. Twisted
projection imaging (TPI) at 3T of a healthy brain. \textbf{B}. Triple-quantum
filtered (TQF)-TPI at 3T of a healthy brain. \textbf{C}. Representative
perfusion weighted imaging (PWI), diffusion weighted imaging (DWI)
at 1.5T and sodium TPI at 4.7T of the brain of a patient with acute
ischemic stroke. These images show the hypoperfused (Tmax+4s) perfusion
maps, the DWI hyperintense (core) in dotted outline and the PWI-DWI
mismatch tissue (penumbra) in solid outline, and corresponding sodium
images, acquired 4h and 25.5h after symptom onset. Figures A and B:
Courtesy of Professor F. Boada, New York University Medical Center
/ University of Pittsburgh Medical Center. Figure C from Tsang A.
et al. Journal of Magnetic Resonance Imaging 33:41 47, 2011. Reproduced
by permission of Wiley-Blackwell.}
\end{figure}

Since the first experiments on sodium MRI, many studies have been
performed on brain, first to show the feasibility of quantitative
brain sodium MRI, and then to evaluate its possible use for assessing
diseases such as tumors, strokes, multiple sclerosis (MS), Alzheimer's
disease (AD) or Huntington's disease (HD), which will be described
in this section. The hypothesis is that sodium MRI could provide direct
non-invasive information on the changes in cellular integrity/viability
through changes in intracellular sodium concentration and/or extracellular
volume due to these pathologies, and complement proton MRI or other
imaging modalities \cite{2005_70,2011_711,1991_6_2,1990_32_5,1985_9_1,1988_30_5}.
Examples of SQ and TQF brain sodium images are shown in Fig. \ref{fig:brain_1}
A-B. Some other examples of sodium images with new TQF schemes can
be found in Ref. \cite{2012_1,2012_67_6,2010_23_10} and with IR in
Ref. \cite{2005_54_5_1305,2011_46_9}.

Thulborn et al. \cite{2009_19_4} proposed to measure tissue sodium
concentration (TSC) and cell volume fraction (CVF) for assessing diseased
tissues in the brain (tumor, stroke), along proton MRI. These two
parameters are termed \textquotedbl{}bioscales\textquotedbl{}, as
they are quantitative parameters with spatial distribution, while
\textquotedbl{}biomarkers\textquotedbl{} are often just binary indicators
of the presence of disease or not. TSC is the volume fraction weighted
mean of the intracellular sodium concentration ({[}Na{]}$_{\mathsf{in}}$
= 10-15 mM) and the extracellular sodium concentration ({[}Na{]}$_{\mathsf{ex}}$
= 140-150 mM) and CVF is around 0.8 in normal brain tissues. Therefore
TSC = CVF$\times${[}Na{]}$_{\mathsf{in}}$ +(1-CVF)$\times${[}Na{]}$_{\mathsf{ex}}$.
After correction of the fractional average water content in normal
brain as around 0.8, the average TSC in brain is calculated as 45-55
mM, which is the range of values usually measured experimentally.
Increasing TSC is supposed to indicate loss of tissue viability while
CVF indicates concomitant decrease in cell density. TSC maps can be
obtained directly from quantitative sodium MRI where reference phantoms
with known sodium concentrations and relaxation times are placed within
the FOV. However CVF cannot be obtained directly, but can be obtained
from TSC through knowledge of {[}Na{]}$_{\mathsf{in}}$ and {[}Na{]}$_{\mathsf{ex}}$,
obtained from TQF sodium MRI for example \cite{2012_NMRbiomed}. 

\subsection*{Strokes}

Stroke is the third largest cause of death in the USA and a leading
cause of long-term disability. Stroke can be classified in two subtypes:
ischemic and hemorrhagic. Most of the strokes are ischemic (around
87 \%). It is very important to intervene as early as possible after
symptom onset in order to reperfuse viable tissues and minimize tissue
loss in order to improve recovery. The usual treatment is trombolysis
by intravenous injection of recombinant tissue plasmogen (tPA) within
a 3-hour window after onset \cite{1995_333_24}, which induce recanalization
of blocked arteries and potentially reperfusion of ischemic tissues.
However, there is a risk also of hemorrhagic transformation or edema
formation due to reperfusion in non-viable tissues which can reverse
the expected outcome of trombolysis. It is therefore important to
assess the viability of tissues and their likelihood to recover before
treatment is applied. 

A combination of proton diffusion weighted imaging (DWI) and perfusion
weighted imaging (PWI) \cite{2001_38_1_19} has been proposed to help
identify patients with more probable improvement in outcome after
trombolysis: ADC maps from DWI can identify regions of cerebral ischemia
(water restriction) within minutes after ischemic onset, reflecting
cytotoxic edema, and PWI can detect regions with perfusion deficits
(regions at risk) within seconds of the ischemic onset. It has been
proposed that the mismatch area between a larger abnormal PWI region
and a smaller lesion in ADC map can represent the ischemic penumbra,
where the tissue is at risk for infarction but still viable, and therefore
could be saved \cite{1999_53_7,1999_30_8,1999_10_3}. However, this
PWI-DWI mismatch method has some serious limitations. It has been
found that DWI does not necessarily represent ischemic tissue damage,
that the limit of salvageable tissue is not limited to the volume
of DWI-PWI mismatch, and that PWI and DWI do not provide information
on the duration of the acute period of ischemia, and cannot establish
the time of symptom onset, therefore unknown-onset patients are excluded
from trombolysis treatment \cite{2011_33_1_41,2012_3_2}. 

A more direct method, such as quantitative sodium MRI, that could
give accurate spatial information on the tissue viability and also
time information about the stroke onset would prove to be very useful
for stoke management. TSC is very sensitive to cell homeostasis and
any loss of cellular energy production will impair the Na$^{+}$/K$^{+}$-ATPase
and induce loss of ion balance across the cell membrane. An increase
in TSC can be associated to increase of intracellular sodium due to
the loss of integrity of the cell, and also increase of extracellular
volume when cells are dying. Studies in animal model and humans have
shown that sodium MRI can measure increases in TSC over the first
few hours and days after induction of cerebral ischemia, and that
sodium MRI could have a potential utility for stroke management \cite{2012_3_2,2009_66_1,1999_213_1,2011_33_1_41,2005_15_3_639}.
These studies have shown that a elevation of 50\% in TSC above the
TSC value in the homologous region in contralateral brain hemisphere
was consistent with completed infarction, which corresponds to a value
of 70 mM in humans. This value could therefore serve as a threshold
for tissue viability and help decision making about which treatment
should be more adapted for the patient. It is also suggested that
the different rates of loss of tissue viability are reflected in the
different rates of change in TSC values between cortex and basal ganglia.
Therefore, clinical decisions to use thrombolytic agents may use different
time windows depending on the location of the stroke. These studies
showed that sodium MRI can be used as a useful complement to DWI and
PWI for managing patients with acute and subacute stroke and that
TSC evolution can help guide trombolysis protocol outside the 3-hour
time window currently used for treatment decision making. 

Some examples of sodium images in human stroke are shown in Fig. \ref{fig:brain_1}
C. In this study, Tsang et al. \cite{2011_33_1_41} showed that sodium
signal intensity cannot be predicted by PWI and that is was not altered
within the PWI-DWI mismatch tissue, irrespective of the interval between
symptom onset and image acquisition, indicating preserved viable tissue
in this region. A combination of DWI, PWI and sodium MRI could therefore
provide useful information on tissue viability in patients with stroke,
and thus despite unknown symptom onset time. Further studies are necessary
to validate all these findings, with optimized sodium acquisitions
with maybe TQF or IR preparation for increasing the weighting of the
images towards intracellular sodium content, but sodium MRI seems
to be very promising for helping stroke management in a quantitative
and non-invasive manner.

\subsection*{Tumors}

Tumor malignancy can be characterized by angiogenesis and cell proliferation
\cite{2000_100_1_57}, among other characteristics. Unregulated cell
division, leading to tumor growth, can be initiated by changes in
Na$^{+}$/H$^{+}$ exchange kinetics and therefore changes in the
intracellular and extracellular pH \cite{1989_49_1}. This mechanism,
associated with reduced Na$^{+}$/K$^{+}$-ATPase activity \cite{1980_255_12}
leads to increased intracellular sodium concentration that can therefore
also be associated with tumor malignancy \cite{1980_40_5,2003_227_2}.
Most likely, the increase in total sodium concentration in malignant
tumors depends on both changes in extracellular volume fraction and
in intracellular sodium content. Similarly, tumor neovascularization
and increase in interstitial space both lead to increased extracellular
volume fraction and are also associated with the potential for tumor
proliferation \cite{1993_10_4_302}. Overall, total sodium concentration
levels in malignant tumors are likely to be elevated, and therefore
could maybe measured by quantitative SQ sodium MRI non-invasively.
Implementing TQF or IR in the sodium acquisition could also provide
some information more specific to changes in the intracellular sodium
content by reducing the weight of fluids (from edema) and/or extracellular
sodium in the contrast of the images. 

The conventional MRI protocol for brain tumor is based on T2 weighted
images and T1 weighted images with and without gadolinium enhancement
for detecting the location and dimension of the tumor, DWI for detecting
the extent of vasogenic edema while excluding cytotoxic edema and
PWI for detecting the regions of the tumor with high vascularity on
the relative cerebral blood volume map, which are consistent with
high-grade tumor \cite{2009_19_4}. But all these changes are generally
late events in tumor onset. Adding sodium MRI to the protocol would
therefore provide direct and more rapid biochemical information on
the tumor metabolism, and also help monitor the immediate effects
and follow up affects over time of cancer therapies.

Ouwerkerk et al. \cite{2003_227_2} combined proton and sodium MRI
at 1.5T to measure the TSC in brain and determine how TSC is altered
in malignant tumors. Sodium images were acquired with a UTE TPI sequence.
Mean TSC (in mmol /kg wet weight) was measured as 60 for grey matter
(GM), 70 for white matter (WM), 135 for cerebrospinal fluid (CSF),
115 for vitreous humor, 100 for tumor, 70 for unaffected contralateral
tissue, and 100 for regions surrounding the tumors (detected with
FLAIR hyperintense proton image). Significant differences in TSC were
demonstrated for both tumors and surrounding FLAIR hyperintense tissues
versus GM, WM, CSF, and contralateral brain tissue. This work shows
that UTE sodium MRI can be used to quantify absolute TSC in patients
with brain tumors and shows increased sodium concentration of 50-60\%
in tumors relative to that in normal brain regions. However, these
measurements cannot define if TSC increases are due to changes in
extracellular volume, intracellular sodium content or neovascularization. 

\begin{figure}[h]
\begin{centering}
\includegraphics[width=1\columnwidth]{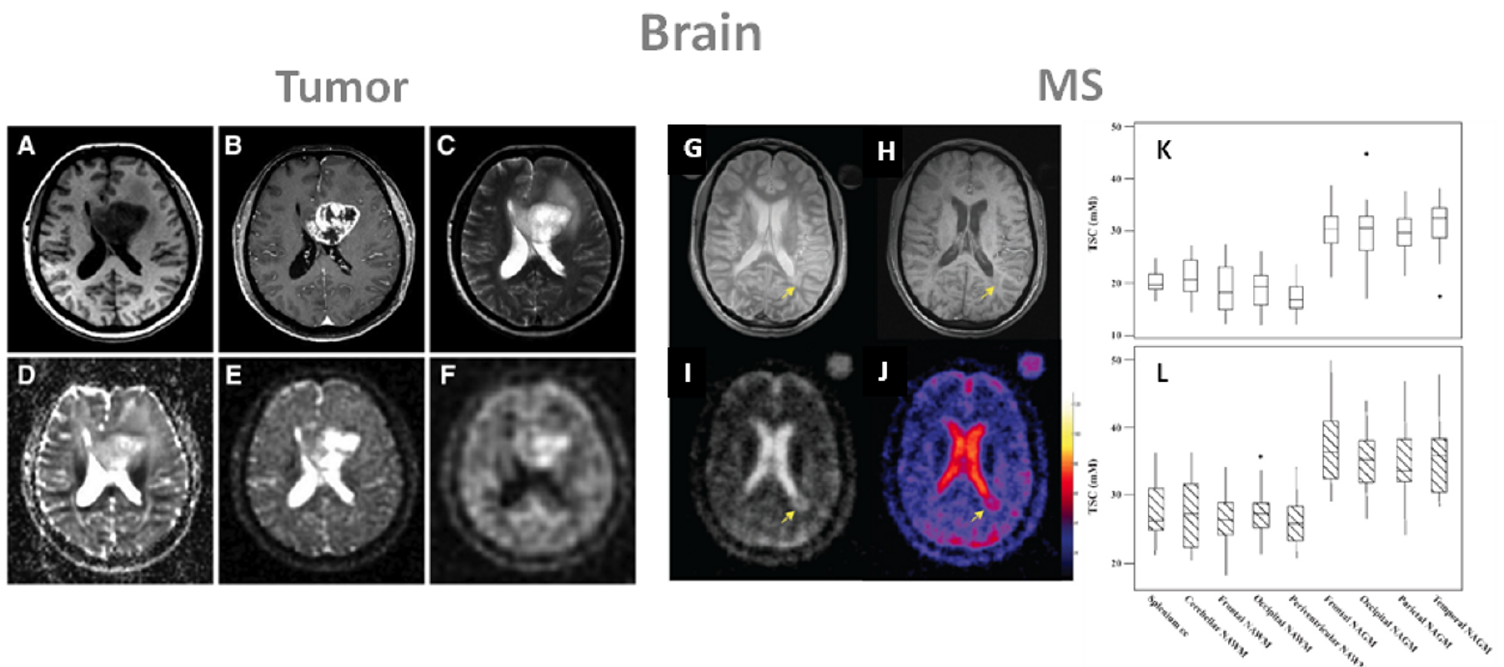}
\par\end{centering}

\caption{\label{fig:brain_2}Examples of brain images. \textbf{A-F}: Proton
and sodium images from a patient with glioblastoma (WHO grade IV)
of the left medial frontal lobe. \textbf{G-L}. Images and tissue sodium
concentrations (TSC) from a patient with multiple sclerosis (MS).\textbf{
A}. T1 weighted MRI, \textbf{B}. T1 weighted MRI with contrast medium
(rim enhancement), \textbf{C}. T2 weighted MRI showing cystic and
solid portions of the lesion and perifocal brain edema, \textbf{D}.
DWI showing elevated ADC values in the center of the tumor, \textbf{E}.
sodium MRI showing increased sodium signal in the tumor, \textbf{F}.
sodium MRI with fluid suppression by inversion recovery (IR), also
showing increased sodium signal mainly at the center of the tumor.
Proton images (\textbf{A-D}) were acquired at 3T while sodium images
(\textbf{E, F}) were acquired at 7T. \textbf{G}. proton density MRI,
\textbf{H}. T1 weighted MRI, \textbf{I}. sodium MRI, \textbf{J}. corresponding
TSC map. \textbf{K}. box-plot of the mean TSC value distribution in
regions of white matter and grey matter in healthy controls. \textbf{L}.
box-plot of the mean TSC value distribution in regions of the corresponding
normal appearing white matter (NAWM) and normal appearing grey matter
(NAGM) in patients with relapsing-remitting MS. Abbreviation: cc is
corpus callosum. Figures A-F from Nagel A. et al. Investigative Radiology
46(9), 539-547, 2011. Reproduced by permission of Wolters Kluwer Health.
Figures G-L from Inglese M. et al. Brain 133, 847-857, 2010. Reproduced
by permission of Oxford University Press.}
\end{figure}

Nagel et al. \cite{2011_46_9} recently applied sodium MRI with and
without IR to 16 patients with brain tumors of different grades (WHO
grades I-IV) at 7T. Some examples of proton and sodium images are
presented in Fig \ref{fig:brain_2} A-F. Their findings were that
TSC imaging revealed increased signal intensities in 15 of 16 brain
tumors before therapy and that IR imaging enabled further differentiation
of these lesions by suppressing CSF and edema signal; all glioblastomas
(grade IV) demonstrated higher IR sodium signal intensities as compared
with WHO grade I-III tumors. It was also noted that contrarily to
total TSC signal, the IR sodium signal correlated with the histologic
MIB-1 proliferation rate of the tumor cells. This study shows that
a combination of sodium MRI with and without T1 relaxation weighting
through IR can reveal important physiological tissue characteristics
and help characterize and grade tumors.

Very recently Fiege et al. \cite{2012_1} combined SQ and TQF sodium
imaging in a single acquisition scheme and applied this new sequence
to 6 healthy brains and 3 brains with tumors at 4T. In this very preliminary
work, they detected a decrease of signal in tumor regions on the TQF
images compared to normal tissue, probably due to suppression of edema
around the tumors. But due to the very low resolution of the images
(10 mm isotropic) and low SNR of the TQF images, it is hard to reach
any definitive conclusion yet.

\subsection*{Multiple Sclerosis}

Another interesting application of sodium MRI in the brain would be
for assessing multiple sclerosis (MS). MS is an inflammatory neurological
disease characterized by focal and diffuse inflammation in white matter
(WM) and grey matter (GM), by demyelination of the axons and by neuro-axonal
injury and loss, but the cellular and molecular mechanisms contributing
to neurodegeneration are still poorly understood \cite{2001_14_3_271}.
Studies have shown that the accumulation of sodium in axons through
non-inactivating sodium channels can promote reverse action of the
Na$^{+}$/Ca$^{2+}$ exchanger, which leads to a metabolic cascade,
and results in an overload of intra-axonal calcium and axon degeneration
\cite{2008_4_3,2004_25_11}.

Inglese et al. \cite{2010_133_3_847} recently demonstrated the first
application of sodium MRI to MS at 3T in 17 patients with relapsing-remitting
MS (rrMS) and 13 normal subjects. The main results are shown in Fig.
\ref{fig:brain_2} G-H. Images were acquired at 3T with a 3D radial
sequence and a birdcage coil. The absolute TSC was measured in lesions
and in several areas of normal-appearing white and grey matter in
patients (NAWM and NAGM in proton MRI), and corresponding areas of
white and grey matter in controls. Mean sodium concentrations were
20 and 30 mM in WM and GM in controls, and 27 and 36 mM in corresponding
NAWM and NAGM in MS patients. In this preliminary study, TSC in MS
patients issue were therefore elevated in acute and chronic lesions
compared to areas of NAWM. The TSC in areas of NAWM were significantly
higher than those in corresponding WM regions in healthy controls.
TSC averaged over lesions and over regions of NAWM and NAGM matter
was positively associated with T2-weighted and T1-weighted lesion
volumes from proton MRI. The expanded disability status scale score
showed a mild, positive association with the mean TSC in chronic lesions,
in regions of NAWM and NAGM. More studies need to be performed to
understand the pathophysiological mechanisms involved in tissue injury
in MS and their link to sodium images. Separation of intracellular
and extracellular sodium by TQF \cite{2012_NMRbiomed} or IR could
prove to be useful for this purpose. But this work shows that abnormal
values of TSC measured non-invasively with sodium MRI in patients
with rrMS might reflect changes in the composition of the lesions
and/or changes in metabolic integrity.

Another recent study by Zaaraoui et al. \cite{2012_264_3} expanded
the method to patients with early and advanced rrMS. It was found
that TSC increased inside demyelinated lesions in both groups of patients.
TSC was also increased in NAMW and NAGM of advanced rrMS patients,
but not in early rrMS.

\subsection*{Alzheimer's Disease}

Finding biomarkers for detecting early signs of AD and tracking response
to treatments is a subject of intense research. Methods such as sampling
the cerebro-spinal fluid (CSF) for biochemical analysis of biomarkers,
positron emission tomography (PET), and MR imaging (through regional
volumetric analysis) or spectroscopy (N-acetylaspartate with 1H NMR
or glutamate with 13C NMR) have been proposed. 

A first study on the applicability of sodium MRI to Alzheimer's disease
(AD) was performed by Mellon et al. \cite{2009_30_5_978}. The hypothesis
here is that alterations of the sodium levels in brain due to cell
death or loss of viability characteristics of AD could be measured
with sodium MRI non-invasively and give useful complementary information
for assessment of early AD. They did not quantify the sodium concentrations
but still were able to detect a small increase (7.5\%) of the sodium
relative signal intensity in the brains of patients with AD compared
to controls and found that this signal intensity enhancement was moderately
inversively correlated with hippocampal volume measured from T1 weighted
inversion recovery proton images. No conclusive explanation on a physiological
basis can be advanced for explaining this sodium content increase.
Is it due to increase of extracellular volume due to cell death and
fluid invasion, increase of intracellular sodium due to an impairment
of Na$^{+}$/K$^{+}$-ATPase, leakage of sodium due to amyloid beta
channels in the membrane, or probably a combination of many of these
causes? More studies that allow fluid suppression and/or intracellular
sodium isolation (IR, TQF) should be performed in order to selectively
study these possible aspects of AD progression.

\subsection*{Huntington's Disease}

A very recent preliminary study \cite{2012_63_1} showed that patients
with Huntington's disease (n=13) also present an increased of TSC
in the whole brain compared to healthy controls (n=13), in structurally
affected regions of the brain, but also in some non-affected regions.
As for the AD study, no satisfying explanation to these TSC increases
could be proposed due to the limited data, low resolution and lack
of differentiation between intracellular and extracellular sodium
content. But further more complete studies may help explain these
observed TSC variations which are generally linked to changes in cellular
and metabolic integrity leading to structural degeneration.

\section{Breast }

Treatments for breast cancer such as prophylactic mastectomy or chemoprevention
are more effective when the disease is detected at an early stage.
The most common techniques for detecting breast cancer are clinical
breast exam (CBE), ultrasound and x-ray mammography. Mammography has
a high sensitivity (70-90\%) but a limited specificity (32-64\%) leading
to many false positives, and very often cannot distinguish benign
from malignant tumors \cite{2009_4}. Moreover, approximately 10\%
of tumors are not detected by mammography due to the presence of dense
fibroglandular tissue. Adding ultrasonography to mammography can help
increasing the sensitivity but also increases the number of false
positives \cite{2008_299_18_2151}. 

\begin{wrapfigure}{o}[30mm]{100mm}%
\begin{centering}
\includegraphics[width=100mm]{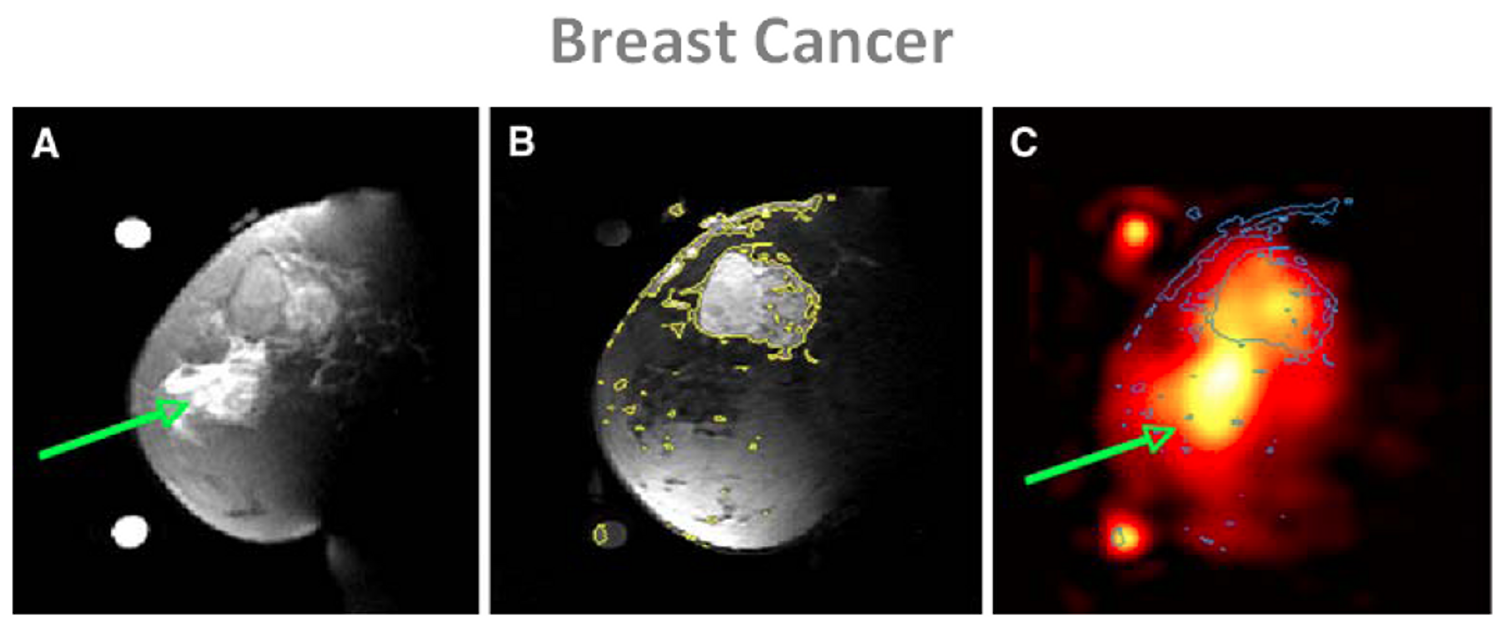}
\par\end{centering}

\caption{\label{fig:breast}Examples of breast cancer images from a patient
with 5.5 cm infiltrating poorly differentiated ductal carcinoma (T3)
at the 12 o'clock position in the left breast. \textbf{A}. Fat suppressed
T2 weighted MRI showing a mass with T2 intermediate signal and edematous
T2 bright retroareolar glandular tissue (arrow). \textbf{B}. Fat suppressed
T1 weighted MRI post-Gd injection with contour levels in yellow, showing
enhancement of the mass at 12 o'clock but not in the retroareolar
glandular tissue. \textbf{C}. Sodium MRI with contours from B superimposed
in blue. Region with edema is indicated by the green arrow. From Ouwerkerk
R. et al. Breast Cancer Res Treat DOI 10.1007/s10549-006-9485-4, 2007.
Reproduced by permission of Springer Science and Business Media.}
\end{wrapfigure}

Proton MRI, particularly contrast-enhanced and diffusion MRI, is a
promising technique for detecting and characterizing tumors not visible
in mammograms, mainly in dense tissues, as MRI has a very high negative
predictive value (rule out the presence of cancer) \cite{2008_148_9_671}.
But the specificity of proton MRI could be improved by adding more
information about the physiology and metabolism of suspicious lesions,
such as cellular integrity and energy metabolism. As proliferating
tumors may cause increases in the sodium content of tissues due to
disruption of the sodium-potassium pump in cell membranes, quantitative
sodium MRI would be a good candidate to detect tumors in the breast
and also assess the degree of malignancy and follow-up chemotherapy.
To test this hypothesis, Ouwerkerk et al. \cite{2007_106_2} applied
quantitative sodium MRI at 1.5T and TPI acquisition to patients with
benign and malignant breast tumors before biopsy. T2 and T1 weighted
contrast-enhanced proton MRI were also acquired. Sodium and proton
images were co-registered to allow quantification of TSC in normal
and suspicious tissues based on $^{\mathsf{1}}$H MRI contrast enhancement,
with histology confirmed by biopsy. The measured TSC were higher by
an average of 60\% in histologically proven malignant lesions compared
to glandular tissue. TSC in benign tumors was significantly higher
than adipose tissue but on the same level than glandular tissue (34
mM). An example of these results in shown in Fig. \ref{fig:breast}.

Increased TSC can arise from increase of extracellular volume fraction
(EVF) due to changes in cellular organization, from increase of vascular
volume, or from increases in the intracellular sodium concentration
due to impaired energy metabolism in Na$^{+}$/K$^{+}$-ATPase activity,
or a combination of all these causes. It is therefore impossible from
this data to reach any conclusion about the measured elevated TSC,
but more studies at higher field, higher resolution, with a combination
of sodium sequences allowing to increase the weight of intracellular
or extracellular sodium on the images (such as TSC or IR), combined
with sensitive proton MRI techniques %
such as DWI or T1 contrast enhanced, may allow to increase the sensitivity
and specificity of MRI in detecting and assessing biochemical information
in breast cancer.

A multiparametric multinuclear combination of MRI techniques could
prove to be useful for grading benign and malignant tumors, and also
for monitoring chemotherapy response \cite{2010_17_12}. It was even
shown by Jacobs et al. \cite{2011_128_1} that a multimodal combination
of $^{\mathsf{1}}$H+$^{\mathsf{23}}$Na MRI with computed tomography
(CT) and proton emission tomography (PET) was feasible and may help
to evaluate the complex tumor micro-environment by examining changes
in morphology, sodium concentrations and glucose metabolism in response
therapy.

\section{Heart}

Acute myocardial infarction (MI) can lead to an increase in the intracellular
sodium concentration due to loss of ionic homeostasis, and to enlarged
extracellular volume due to myocardial edema formation or due to scar
formation \cite{2004_110_22,2001_45_5}. Quantitative sodium MRI seems
therefore a good candidate to try to detect cardiac infarction by
measuring localized increased sodium content in cardiac tissues and
help differentiate viable from non-viable tissues. Some preliminaries
studies using surface coils were performed over the years to test
the feasibility of cardiac sodium MRI and to try to assess infarction
\cite{2004_16_6,1997_38_4,2005_21_5_546,2008_248_1,2001_45_1_164}. 

Ouwerkerk et al. \cite{2005_21_5_546} measured the myocardial TSC
of healthy volunteers. Mean TSC was approximately 43 mM in the left
ventricular (LV) free wall, 53 mM in the septum and 17 mM in adipose
tissue. The same team then applied the method to twenty patients with
nonacute MI. Mean TSC for MI was measured as 30\% higher than noninfarcted
tissues in LV regions (significant difference). The mean TSC in regions
adjacent to MI regions was intermediate between MI and normal tissue
sodium content. They also conclude that TSC increase is not related
to infarct age, size or global ventricular function. The measured
sodium TSC or pixel intensity changes may be attributable to loss
of cellular integrity, inhibition of the Na$^{+}$/K$^{+}$-ATPase
function due to energy depletion and changes in the sodium concentration
gradient between intracellular and extracellular volumes, and also
possibly changes in the sodium ion molecular environment, and probably
a combination of all. 

Because of the methods used until now, it is difficult to find in
which proportion and when these probable physiological changes can
be associated with observable changes in sodium content or relaxation.
Future studies might include imaging at higher fields for increasing
the SNR and spatial resolution for better identification of infarcted
tissues and adjacent areas, and sequences that could separate intracellular
from extracellular sodium and/or relaxation times weighted (TPI associated
with TQF or IR), for a better estimation of intracellular TSC, which
is generally supposed to be more sensitive to cell energy impairment
and viability.

\section{Muscle}

Sodium MRI has the potential to provide insights into muscle physiology
and for assessing muscle diseases. In muscle tissues, the electrochemical
gradient across the cell membrane is maintained by the Na$^{+}$/K$^{+}$-ATPase,
but when an action potential leading to muscle contraction is generated,
there is a rapid influx of sodium ions and efflux of potassium ions
via the sodium and potassium channels. During intense contractile
activity, the persistent influx and efflux of ions, which degrades
the transmembrane Na$^{+}$ and K$^{+}$ gradients, can lead to a
loss of membrane excitability and muscle contractility \cite{2003_83_4},
which is believed to represent one of the main mechanisms of muscle
fatigue \cite{2008_104_1}.

It has already been shown that many disease states, such as diabetes,
starvation and hypothyroidism, can be linked to a decrease in Na$^{+}$/K$^{+}$-ATPase
activity in skeletal muscle \cite{2003_83_4}. Therefore sodium MRI
has the potential to play a role in imaging of these disorders. For
example, Bansal et al. \cite{2000_11_5} studied the change of sodium
concentration and relaxation in muscle after voluntary muscle contractions.
They showed that the sodium intensity in the images increased by around
34\% in the exercised muscle and then diminished with a half-time
of 30 min, but that the calculated sodium concentration did not change
significantly, while the long T2 component of the sodium relaxation
increased. They therefore suggest that the change in intensity in
the sodium images was mainly due to a change in the sodium-macromolecules
interaction rather than a change in TSC. We present in the following
paragraphs a few examples of studies of sodium MRI for different muscular
diseases.

\subsection*{Diabetes}

\begin{wrapfigure}{o}[30mm]{100mm}%
\begin{centering}
\vspace{-10mm}
\par\end{centering}

\begin{centering}
\includegraphics[width=100mm]{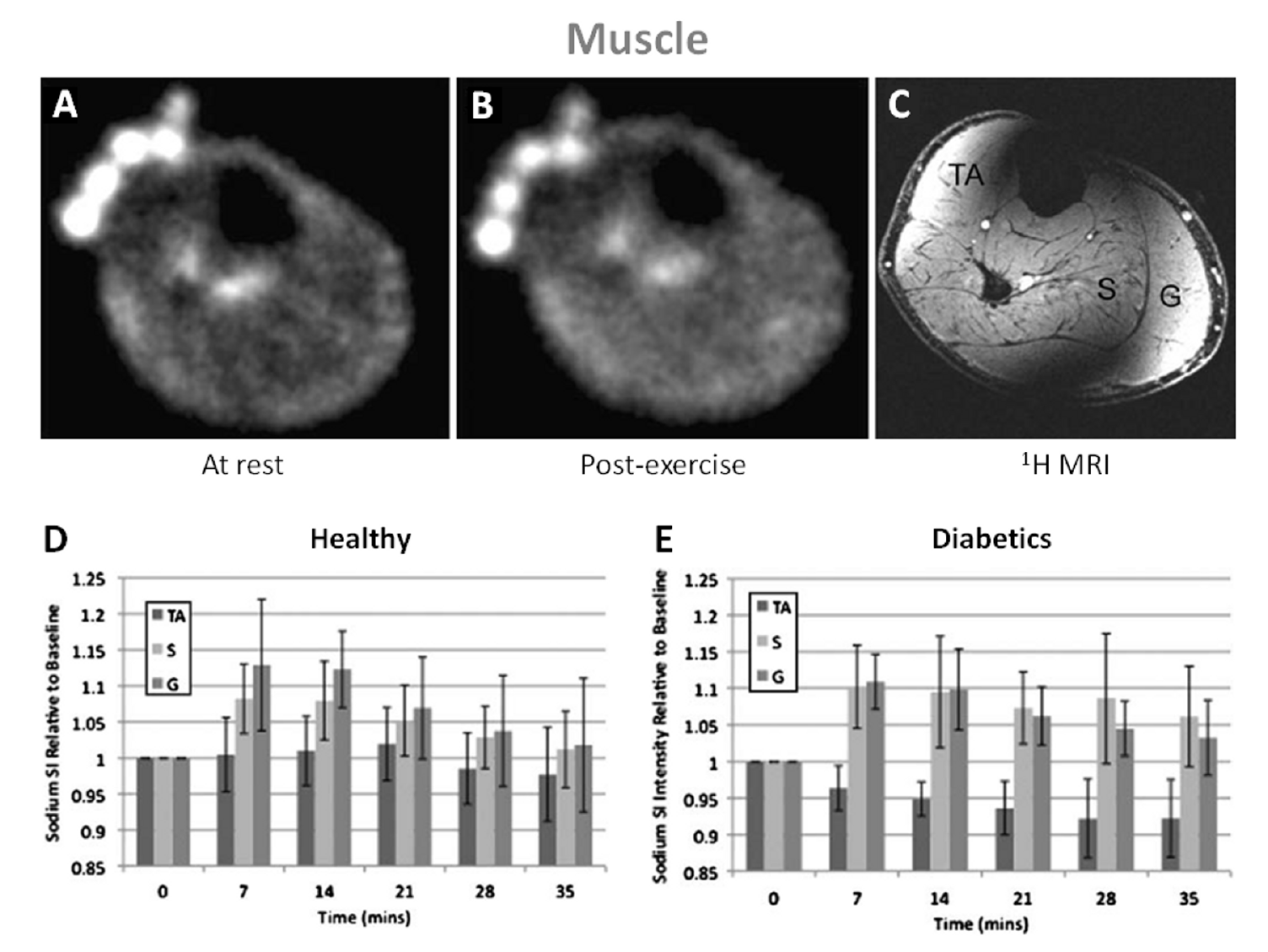}
\par\end{centering}

\caption{\label{fig:muscle}Example of sodium images in the muscle at rest
(\textbf{A}) and post-exercise (\textbf{B}). 3D FLASH proton MRI (\textbf{C})
was performed after the sodium MRI to delineate muscle anatomy: TA
is tibialis anterior, S is soleus and G is gastrocnemius. In healthy
subjects (\textbf{D}), sodium signal intensity increases significantly
in S and G just after exercice, but not in the control muscle TA.
The sodium intensity then decreases to near baseline in both S and
G with a half-time of around 22 min. In diabetics subjects (\textbf{E}),
sodium signal intensity increases significantly in S and G just after
exercice, but not in the control muscle TA. The sodium intensity then
decreases to near baseline half-times of around 37 min in S and 27
min in G. From Chang G. et al. European Radiology, DOI 10.1007/s00330-010-1761-3,
2010. Reproduced by permission of Springer Science and Business Media. }
\end{wrapfigure}

Sodium MRI has been applied recently on diabetic patients. Some results
from the study by Chang et al. \cite{2010_20_8} are presented in
Fig. \ref{fig:muscle}. Pre- and post-exercise sodium intensity (SI)
was measured in sodium images in healthy volunteers and patients with
diabetes, in the tibialis anterior (TA) as control muscle, in soleus
(S) and in gastrocnemius (G) muscles. It was found that the muscle
sodium signal intensities (in S and G) increase significantly immediately
after exercise and that afterwards this sodium signal recovers down
to baseline more slowly in diabetics than in healthy subjects.

The authors suggest that the biophysical basis for the increase in
sodium signal intensity in muscle after exercise is probably multifactorial
and could be due mainly to 2 main factors: (1) an increase in total
muscle sodium content due to increase in volumes of the intracellular
and extracellular compartments after exercise and to the increase
of intracellular concentration muscle cell depolarization, therefore,
the increase in sodium signal intensity following exercise could reflect
changes in the quantity of sodium in both compartments; (2) alterations
in sodium macromolecular environment and therefore changes in T2 relaxation
times or in the proportions of the long and short T2 components. One
possible explanation might be the decreased Na$^{+}$/K$^{+}$-ATPase
activity as well as the altered tissue microvasculature in diabetics.
Na$^{+}$/K$^{+}$-ATPase preserves the muscle membrane excitability
by maintaining the Na$^{+}$ and K$^{+}$ concentration gradients
across the cell membrane and therefore helps protect muscles against
fatigue \cite{2003_83_4}. However, diabetics have a decreased Na$^{+}$/K$^{+}$-ATPase
activity and decreased numbers of Na$^{+}$ and K$^{+}$ pumps on
the muscle cell membrane, which has been attributed to insulin resistance
\cite{2001_47_2}. This would result in a decreased ability to extrude
intracellular sodium ions into the extracellular space and result
in elevated muscle sodium intracellular content with a slower recovery
to baseline. 

This preliminary study shows that sodium MRI could therefore be applied
to patients with diabetes in order to evaluate those who are at risk
of diabetic muscle infarction, which is a grave complication of longstanding
diabetes. Moreover, fluid and/or extracellular sodium suppression
by TQF or IR, along with T2 relaxation measurements, could prove to
be useful to assess the origin of the SI decrease observed in this
study.

\subsection*{Muscular Channelopathy}

Patients with hypokalemic periodic paralysis (hypoPP) or paramyotonia
congenita (PC), two different kinds of muscular channelopathies, were
scanned with sodium MRI in a study by Nagel et al. \cite{2011_46_12}.
These rare diseases are considered to be caused by genetic mutations
of the voltage-gated sodium channels in muscular cells, and that can
be characterized by elevated myoplasmic sodium at rest and after cooling
(for provocating PC pathology effects). 

Three sodium techniques were used to assess the disease: total tissue
sodium concentration ($^{\mathsf{23}}$Na-TSC), T1 weighted sodium
imaging ($^{\mathsf{23}}$Na-T1) and inversion recovery ($^{\mathsf{23}}$Na-IR)
which was used to suppress most of the extracellular and vascular
sodium and vasogenic edema. All $^{\mathsf{23}}$Na sequences showed
significantly higher signal intensities in hypoPP compared with PC
patients and healthy subjects, and provocation in PC induced a significant
increase (>20\%) in the $^{\mathsf{23}}$Na-IR signal and a corresponding
decrease of muscle strength. Signal intensities from $^{\mathsf{23}}$Na-T1
and $^{\mathsf{23}}$Na-TSC also have a non-significant tendency to
be higher after provocation in PC. This study indicates that $^{\mathsf{23}}$Na-IR
could provide a stronger weighting toward intracellular sodium than
$^{\mathsf{23}}$Na-T1 or $^{\mathsf{23}}$Na-TSC, and therefore may
allow a better visualization of changes in intracellular sodium content.
A combined application of $^{\mathsf{23}}$Na-TSC and $^{\mathsf{23}}$Na-IR
could therefore improve the analysis of pathophysiological changes
in muscles of patients with muscular channelopathies by measuring
the changes in intracellular concentration and T1 relaxation time. 

The same kind of study was also performed on patient with hyperkalemic
periodic paralysis (hyperPP) by Armateifo et al. \cite{2012_264_1}.
They showed that sodium MRI can detect increased myoplasmic sodium
content in HyperPP patients with permanent weakness, as they are affected
by an incomplete inactivation of muscular sodium channels \cite{2001_35_2}.
In this case too, $^{\mathsf{23}}$Na-IR is more sensitive to intracellular
changes than $^{\mathsf{23}}$Na-TSC and $^{\mathsf{23}}$Na-T1. In
conclusion, sodium overload may cause muscle degeneration developing
with age and sodium MRI could therefore help monitoring medical treatments
that reduce this overload.

\subsection*{Myotionic Dystrophy}

Myotonic dystrophy has been linked to alterations in sodium channel
conductance regulation, which can cause elevated muscle fiber concentrations
that correlate with disease severity \cite{1997_37_2}. 

Constantinides et al. \cite{2000_216_2_559} have found in their study
that the mean TSC measured with sodium MRI after exercise were elevated
by 16\% and 22\% in two healthy volunteers, and 47\% and 70\% in two
dystrophic muscles in compared with those at normal resting levels.
These results in patients with myotonic dystrophy are consistent with
the known imbalance in sodium homeostasis in dystrophic muscle fibers
\cite{1968_214_2}. Quantitative sodium imaging could be a valuable
tool for characterizing the early onset, pathogenesis, and monitoring
of pharmacologic treatment of dystrophic muscle, but more data need
to be acquired to confirm these findings.

\subsection*{Hypertension}

Another interesting possible application of quantitative sodium MRI
in muscle is to measure the increase of body sodium content due to
hypertension, which is linked to a disturbed total body sodium regulation.

Kopp et al. \cite{2012_59_1} measured sodium content in triceps surea
of healthy volunteers and of patients with primary aldosteronism,
before and after treatment. They found a 29\% increase in muscle Na$^{+}$
content in patients with aldosteronism compared with normal women
and men. This tissue Na$^{+}$ content was then reduced to normal
levels (20-25 mM) after successful treatment without accompanying
weight loss. They suggest that sites such as muscle (and also skin)
could serve to store Na$^{+}$ nonosmotically by binding of the Na$^{+}$
ion to proteoglycans within the extracellular compartment without
apparent accompanying fluid retention or changes in serum Na+ concentration
in patients with primary aldosteronism. 

Sodium MRI in muscle may therefore be useful for detecting and follow
up treatment of hypertension and also help testing the role of Na$^{+}$
for assessing long-term cardiovascular risk in populations.

\section{Cartilage}

Cartilage is a dense connective tissue that can be found in many parts
of the body such as articular joints between bones (hyaline cartilage),
in the ear and nose (elastic cartilage) or in intervertebral disk
(fibrocartilage). 

In this review article we will mainly focus on articular hyaline cartilage,
which consists of a small population of chondrocytes (5\% of volume)
within a large extracellular matrix (ECM) made of type II collagen
fibers (15 20\% of the volume), proteoglycans (PG; 3 10\%) and water
(65 80\%) and does not contain blood vessels. PGs further consist
of a protein core and negatively charged glycosaminoglycan (GAG) side
chains, which endow the cartilage with a negative fixed charge density
(FCD). This FCD attracts free-floating positive counter-ions, such
as sodium $^{\mathsf{23}}$Na$^{+}$, which in turn attract water
molecules within the cartilage through osmotic pressure. Therefore,
we can notice that unlike most of biological tissues, sodium ions
in cartilage are mostly present in the extracellular volume. The negative
charge from the GAG side chains also provides a strong electrostatic
repulsive force between the PG molecules and is responsible for the
compressive stiffness of cartilage. The collagen fibers serve to immobilize
the PG and provide a tensile force opposing the tendency of the PG
to expand the cartilage. Thanks to these properties, articular cartilage
can provide synovial joints with lubrication and also serve to absorb
mechanical shocks and to distribute load over the underlying bone,
and thus makes normal motion of possible \cite{2006_19_7,1994_1_44}.

\subsection*{Osteoarthritis}

Osteoarthritis (OA) is the most common form of arthritis in synovial
joints and a leading cause of chronic disability, mainly in the elderly
population. From the biochemical point of view, OA is a degenerative
disease of the articular cartilage mainly characterized by a reduction
of FCD (or GAG) concentration, possible changes of size and organization
of the collagen fibers, aggregation of the PG and increased water
content. These changes lead to an alteration of the mechanical properties
of the cartilage, which can therefore lose its load- and shear-bearing
functions. At present, there is neither a known cure nor a preventive
treatment for OA, and present treatments focus mainly on pain management
with analgesics and improvement in quality of life (e.g., exercise
and weight loss). If these methods are ineffective, joint replacement
surgery may be considered. Early detection of OA, prior to irreversible
morphological changes, and an accurate method for quantifying the
effects of potential treatments are therefore of fundamental importance. 

\begin{wrapfigure}{o}[30mm]{65mm}%
\begin{centering}
\vspace{-4mm}
\par\end{centering}

\begin{centering}
\includegraphics[width=65mm]{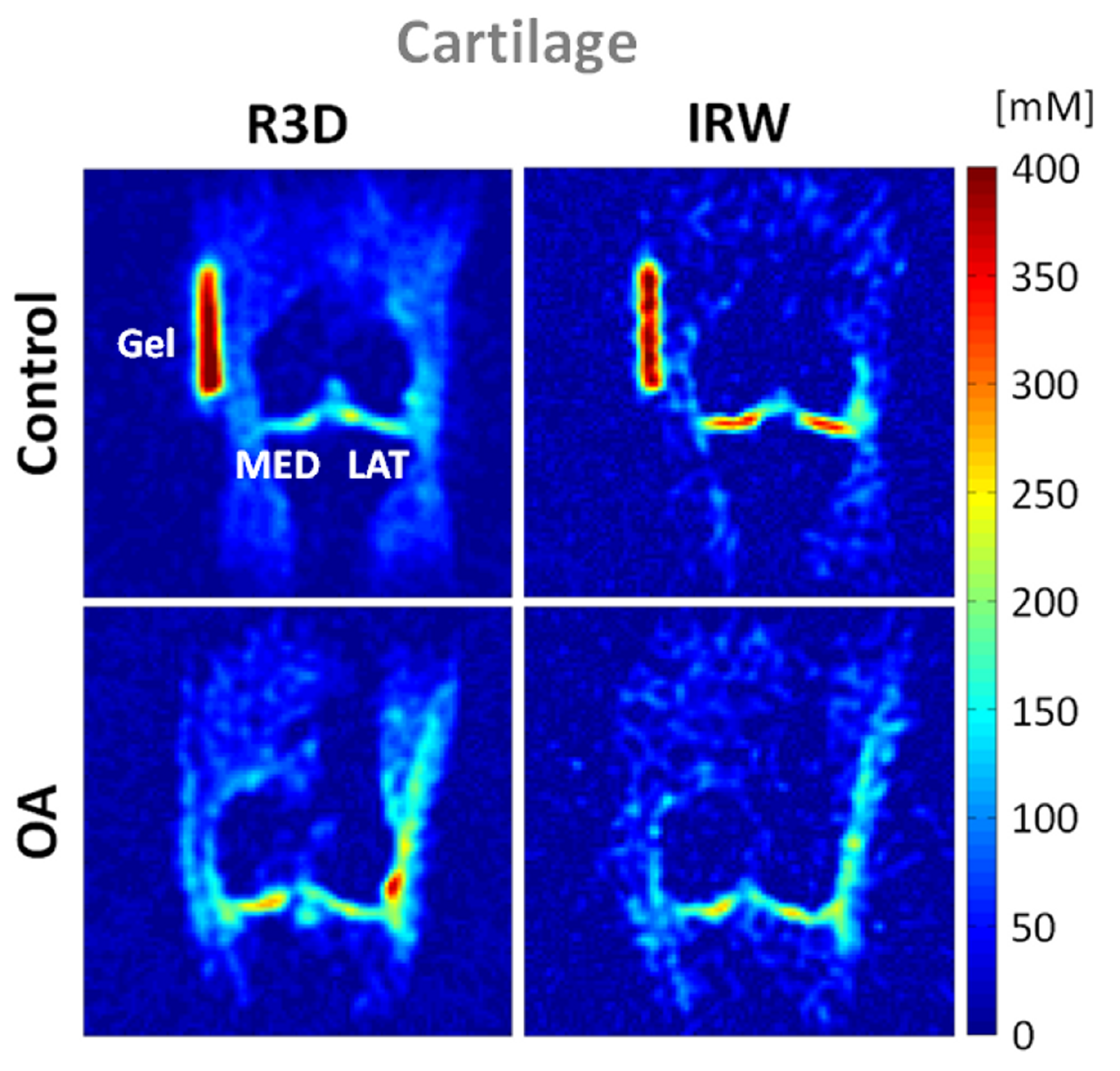}
\par\end{centering}

\caption{\label{fig:cartilage}Cartilage sodium concentration maps from a healthy
volunteer (control) and a patient with osteoarthritis (OA). Images
were acquired with 3D radial (R3D) and IR WURST (IRW). IR WURST was
used to suppress fluids through inversion recovery (IR) with an adiabatic
WURST pulse in order to increase the sensitivity of the method to
change of sodium concentration within the cartilage only. MED = femoro-tibial
medial and LAT = femoro-tibial lateral cartilage.}
\end{wrapfigure}%

Radiography is the standard method used to detect gross loss of cartilage
by measuring narrowing of the distance between the two adjacent bones
of a joint, but does not image cartilage directly. Proton MRI, such
as T1-weighted, T2-weighted and proton density, can provide morphological
information on damage of cartilage, such as fissuring and partial-
or full-thickness cartilage defects, but does not give any information
on the GAG content within the ECM or the structure of the collagen
fibers network. New methods for functional proton MRI of the cartilage
are under now development, such as T1$\rho$ mapping \cite{2003_229_1},
T2 mapping \cite{2004_8_4}, GAG chemical exchange saturation transfer
(gagCEST) \cite{2008_105_7}, delayed gadolinium enhanced MRI of cartilage
(dGEMRIC) \cite{1996_36_5} and diffusion weighted imaging (DWI) \cite{2005_53_5}.
T2 mapping and DWI can give information about the collagen fibers
and water mobility, while the other methods can give some information
about the FCD or GAG content, indirectly through the use of a contrast
agent (dGEMRIC), or directly (T1$\rho$ mapping, gagCEST). All these
methods are still under investigation for assessing their specificities
and sensitivities for detecting early OA. 

It has been shown that sodium concentration has a strong correlation
with FCD and GAG content in cartilage \cite{1992_10_1}, therefore
quantitative sodium MRI could also be useful for detecting directly
loss of GAG in early OA \cite{1998_39_5,2002_47_2_284,2004_231_3}.
Several studies have already been performed on healthy and OA cartilage,
and show that in general TSC in healthy cartilage is in the range
250-350 mM, while <250 mM in OA cartilage \cite{2004_231_3}. Because
of the low resolution of the sodium images (>3 mm), due to the presence
of synovial fluid or joint effusion and also possible cartilage thickening,
the sensitivity of the method to measure small changes of TSC within
the cartilage only should include fluid suppression by either TQF
\cite{1997_38_2}\cite{1997_38_2,1999_141_2} or IR \cite{2008_193_2,2010_207_1}. 

An example of quantitative sodium MRI of cartilage at 7T in presented
in Fig. \ref{fig:cartilage}, where fluid suppression was obtained
by IR with an adiabatic inversion pulse \cite{2010_207_1}. Images
were acquired with a 3D radial sequence. It is shown that fluid suppression
allows a better differentiation between control and OA patient. This
work is still preliminary but it is reproducible and repeatable \cite{2011_repeat}.
Another possible application under investigation would be the measurements
of the relaxation times of sodium in cartilage, which are expected
to be very sensitive to any change of their environment, such has
GAG depletion of collagen fibers rupture. Some reviews on the potential
of sodium MRI and OA are presented in Ref. \cite{2006_19_7,2011_3_1_1}.

\subsection*{Cartilage Repair}

Sodium MRI could also prove to be useful for assessing cartilage repair.
Trattnig et al. \cite{2010_257_1} showed the feasibility of dGEMRIC
and sodium MRI at 7T for differentiating repaired tissue after matrix-associated
autologous chondrocyte transplantation (MACT) from native cartilage
at 3T. A strong correlation was found between sodium imaging and dGEMRIC
in patients after MACT. Another recent study by Chang et al. \cite{2012_22_6}
showed that the method for assessing sodium concentration in cartilage
repair can be improved by fluid suppression by IR.

\subsection*{Intervertebral Disk}

Insko et al. \cite{2002_9_7} demonstrated the feasibility of sodium
MRI of the intervertebral disk (IVD) \emph{in vivo} at 4T using a
surface coil and 3D gradient echo sequence. This technique could be
used to measure the proteoglycan content in fibrocartilage and help
detect early degenerative changes in IVD. In a subsequent recent study,
Wang et al. \cite{2010_35_5_505}, showed that sodium MRI can correlate
with loss of proteoglycans in the spine.

\section{Abdomen}

Quantitative sodium MRI could also prove to be useful for assessing
cell viability in abdominal organs and in order to detect and diagnose
diseases in liver, gallbladder, pancreas, kidney, spleen, prostate,
uterus or other organs \cite{2009_13_138}. Very few studies have
been performed for the moment on the abdomen, mainly due to the lack
of availability of body sodium coils and the necessity for the acquisition
sequences to take into account the cardiac and respiratoty movements
of the body, which can seriously perturb the quality of the images.
However, some preliminary work is presented here on kidney, prostate
and uterus.

\subsection*{Kidney}

The kidney is essential in regulating homeostatic functions in the
body such as extracellular fluid volume, acid-base balance (pH), electrolyte
concentrations, and blood pressure (via maintaining salt and water
balance). This role depends tightly on the regulation of extracellular
sodium in the kidney, which builds up a concentration gradient from
the cortex to the medulla. Thus renal function is tightly dependent
on this corticomedullary gradient and mapping this gradient with sodium
MRI could help assess kidney impairments. 

\begin{wrapfigure}{o}[30mm]{100mm}%
\begin{centering}
\vspace{-2mm}
\par\end{centering}

\begin{centering}
\includegraphics[width=100mm]{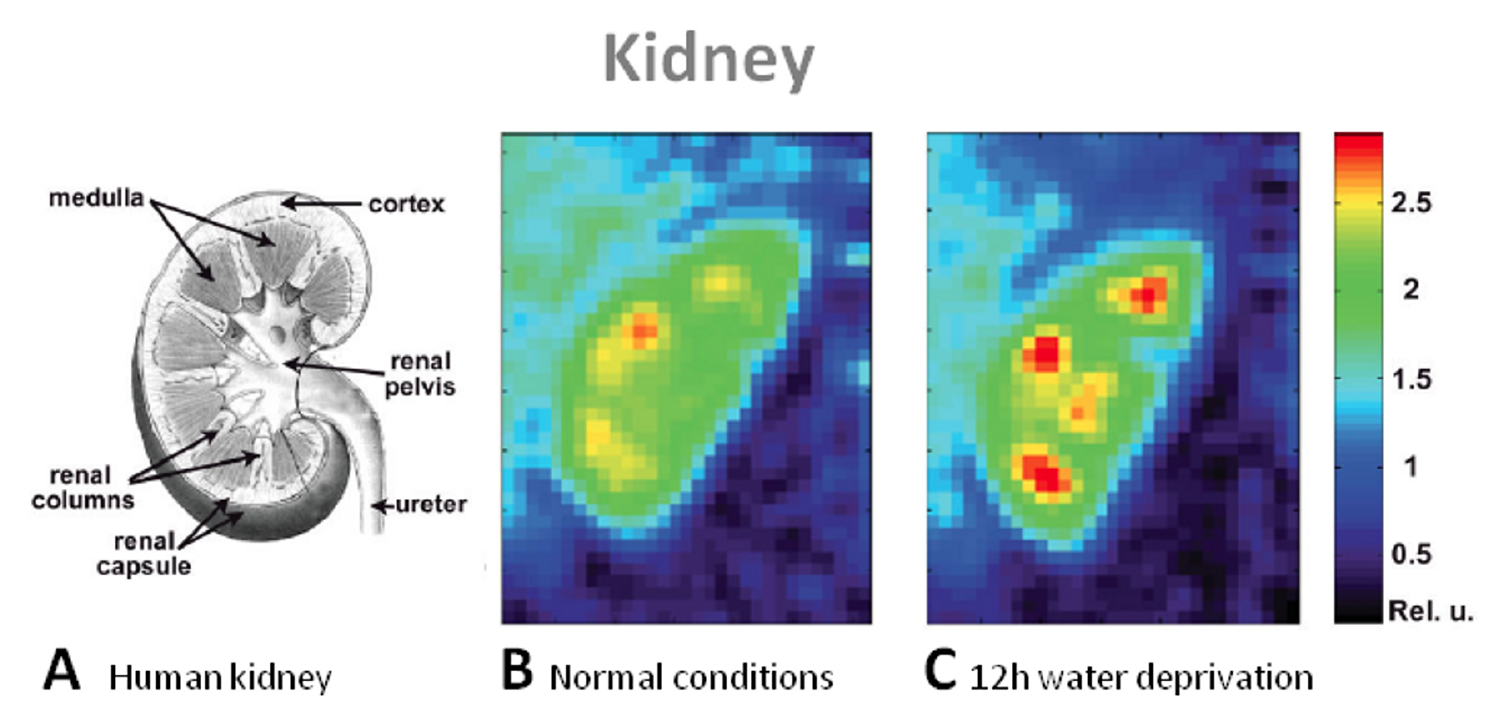}
\par\end{centering}

\caption{\label{fig:kidney}Examples of kidney images. \textbf{A}. Scheme of
a human kidney. Central coronal slices of the 3D sodium images of
a human kidney under normal conditions (\textbf{B}) and 12-h water
deprivation (\textbf{C}). The sodium gradient increases significantly
by 25\% after water deprivation. From Maril N. et al. Magnetic Resonance
in Medicine 56, 1229 1234, 2006. Reproduced by permission of Wiley-Blackwell.}
\end{wrapfigure}%

A first study in human by Maril et al. \cite{2006_56_6} presents
sodium MRI of kidneys before after water deprivation, as shown in
Fig. \ref{fig:kidney}. This work was performed at 3T with a surface
coil and data was acquired with a 3D gradient echo sequence. The results
show that the sodium signal intensity increased linearly from the
cortex to each of the medullae with a mean slope of 1.6 (in arbitrary
units per mm) and then decreased toward the renal pelvis. After 12
hour water deprivation, this gradient increased significantly by 25\%.
The sodium gradient change in the kidney may reflect changes in the
concentration in each kidney compartment (cortex, medulla), but also
micro-compartments within these compartments (intracellular, extracellular,
vascular), differences in the relaxation times of sodium within the
tissues, or a combination of all of them.

A following study by Rosen et al. \cite{2009_16_7} performed sodium
MRI at 3T on a patient with a transplanted kidney, who was previously
diagnosed with end-stage hypertensive nephropathy, to measure the
corticomedullary sodium gradient and to assess the good functioning
of the new kidney. The kidney was imaged 4 months after transplantation.
The measured mean medulla/cortex SNR ratio was of 1.8, and the gradient
slope was 1.1 (arbitrary units / mm), which is lower than the gradient
observed in healthy kidney in the previous study. A interpretation
can be made at this stage as the kidney was recently transplanted,
resolution and SNR were low and the physiology/biochemistry involved
are not fully known. More studies need to be done of course, but these
findings are encouraging for a possible application of quantitative
sodium MRI for assessing renal functions in different diseases such
as nephropathy, renal failure and kidney transplantation.

\subsection*{Prostate}

Sodium MRI of the human prostate was recently tested \emph{in vivo}
at 3T \cite{2012_47_12,2012_ISMRM_546} and compared with diffusion
MRI. The prostate and its different departments were identifiable,
with measured TSC of around 60 mM in the central zone and 70 mM in
the peripheral zone. This method could be a potential radiological
biomarker of prostate cancer and of treatment response.

\subsection*{Uterus}

Uterine leiomyomata (fibroids) were investigated by diffusion weighted
imaging (DWI) and quantitative sodium MRI by Jacobs et al. \cite{2009_29_3_649}
at 1.5T. Uterine leiomyomata are solid masses arising from the muscle
of the uterus (myometrium) and can be associated with menstrual pain
and loss of reproductive function. The goal was to monitor the treatment
of fibroids on patients that were treated using MRI guided high-intensity
focused ultrasound surgery (MRg-HIFUS). Regions where the tissue was
treated were clearly identified on both DWI and sodium images. The
TSC in normal myometrium tissue was approximately 36 mM with apparent
diffusion coefficient (ADC) of 2.2 mm2/s. The TSC was 28 mM in untreated
fibroids and increased to 42 mM in treated tissues while ADC was 1.75
mm2/s in untreated fibroids and decreased to 1.3 mm2/s in treated
tissues. 

The mechanisms involved in the changes in ADC and TSC in treated uterine
tissue is still unknown but these changes may provide an empirical
measure of the efficacy of the treatment. DWI is sensitive to translational
motion and changes in the intra- and extracellular compartments available
for diffusion of water molecules and can reveal disruptions or restrictions
to the movement of these molecules within a tissue. Sodium MRI is
more sensitive to the gradient of ion concentration between the intra-
and extracellular compartments, which depends on the Na$^{+}$/K$^{+}$-ATPase
activity within the cell membrane and its energy consumption. Thermal
ablation disrupts the cell membrane and alters the cellular integrity
and perfusion of the tissue. The ADC and sodium changes within the
diseased uterine tissue before and after treatment may therefore reflect
a combination of complex changes in tissues such as modification of
the water environment, disruption of Na$^{+}$/K$^{+}$-ATPase function,
decreased vascularization or cytotoxic edema. More data must be acquired
to fully understand these findings, but this preliminary study suggests
that DWI (ADC map) coupled with quantitative sodium concentrations
may provide useful non-invasive biomarkers for \begin{wrapfigure}{i}[10mm]{55mm}%
\begin{centering}
\vspace{-6mm}
\par\end{centering}

\begin{centering}
\includegraphics[width=55mm]{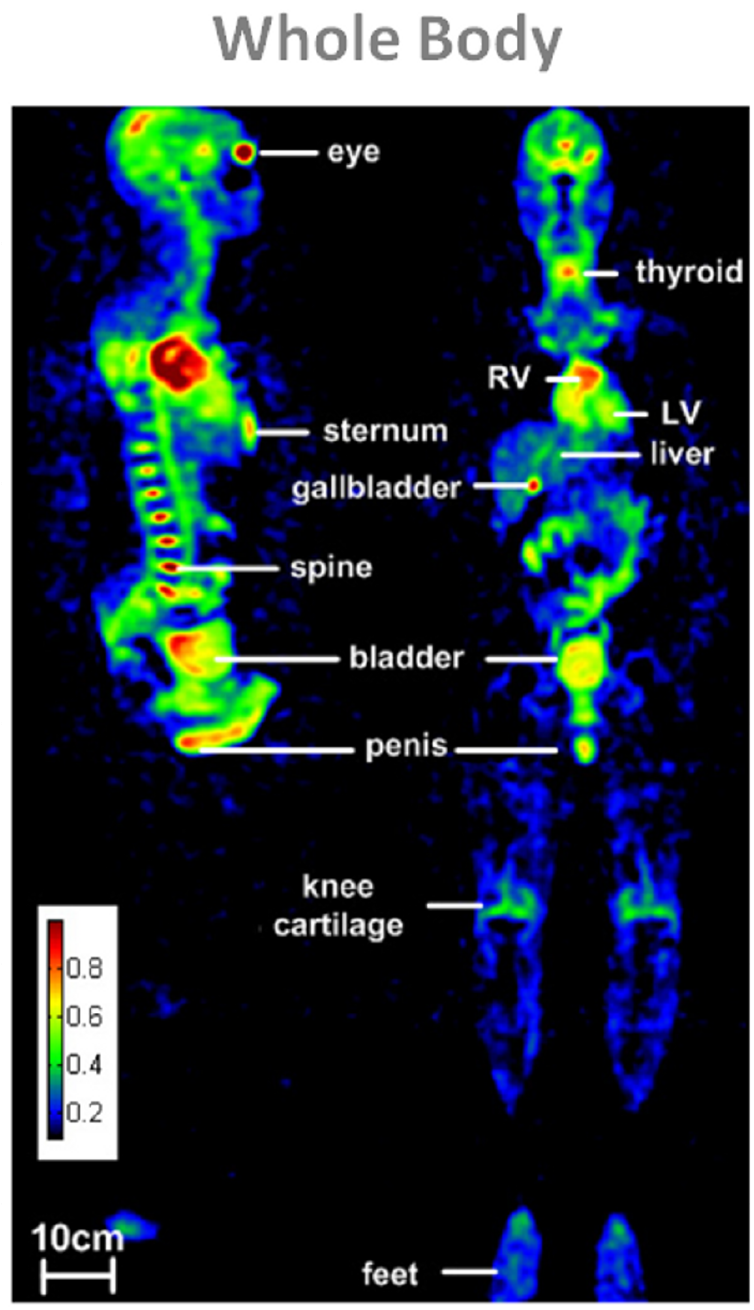}
\par\end{centering}

\caption{\label{fig:wholebody}Whole body sodium MRI of a volunteer. From Wetterling
N. et al. Phys. Med. Biol. 57, 4555-4567, 2012. Reproduced by permission
of IOP Publishing Ltd.}

\centering{}\vspace{-10mm}\end{wrapfigure}%
 uterine leiomyomata response to therapy or to investigate malignant
tumors, by probing biochemical changes that can give information beyond
the usual anatomical imaging parameters.

\section{Whole Body }

Most of human parts have been scanned with sodium MRI separately for
the moment, and recently a study by Wetterling et al. \cite{2012_57_14}
showed a whole body sodium MRI \emph{in vivo} at 3T. A new asymmetrical
birdcage coil was used to acquired the data in 5 segments of 10 min
each with a 3D radial sequence, and a nominal isotropic resolution
of 6 mm. The resulting composite image of the 5 acquisitions is shown
in Fig. \ref{fig:wholebody}. Despite the low resolution, we can easily
detect many different organs such as the brain, the left and right
ventricles of the heart, the liver, the bladder, the spine or articular
cartilage. A possible application of whole body sodium imaging would
be for example the detection and malignancy assessment of cancer with
metastases, and follow-up of the chemotherapeutic response.

\section{Cancer and Chemotherapeutic Response}

Although the following studies were not performed on humans but on
rats or mice \emph{in vivo}, it is useful to mention them as they
represent an important preliminary work previous to potential human
applications on cancer and chemotherapeutic response assessment \cite{2000_6_6,2007_25_7,2005_53_1}.
The goal of cancer therapy is to eliminate neoplastic cells, but assessment
of the therapeutic efficiency \emph{in vivo} may be difficult. Different
functional imaging modalities such as PET (rate of glycolysis), DWI
MRI (extracellular volume) or sodium MRI (intracellular sodium) could
prove to be useful for this purpose.

\subsection*{Chemotherapy Assessment}

It has already been demonstrated that DWI can be useful for quantitative
assessment of tumor response to therapeutic insult by observing the
microscopic changes in water diffusion due to cellular destruction
and that is highly sensitive to microscopic changes in cellular structure
\cite{2002_1_4}. These microscopic changes generally precede macroscopic
changes in tumor volume, which occur at much later stages after therapeutic
insult. On the other hand, quantitative sodium MRI can give direct
information of the intracellular and extracellular sodium concentrations,
which depend on the good function of the sodium channels and Na$^{+}$/K$^{+}$-ATPase
in the cell membrane and cellular energy consumption. Therefore any
disruption in cellular integrity can lead to abnormal imbalances that
are detectable using sodium MRI. 

\begin{figure}[h]
\begin{centering}
\includegraphics[width=1\columnwidth]{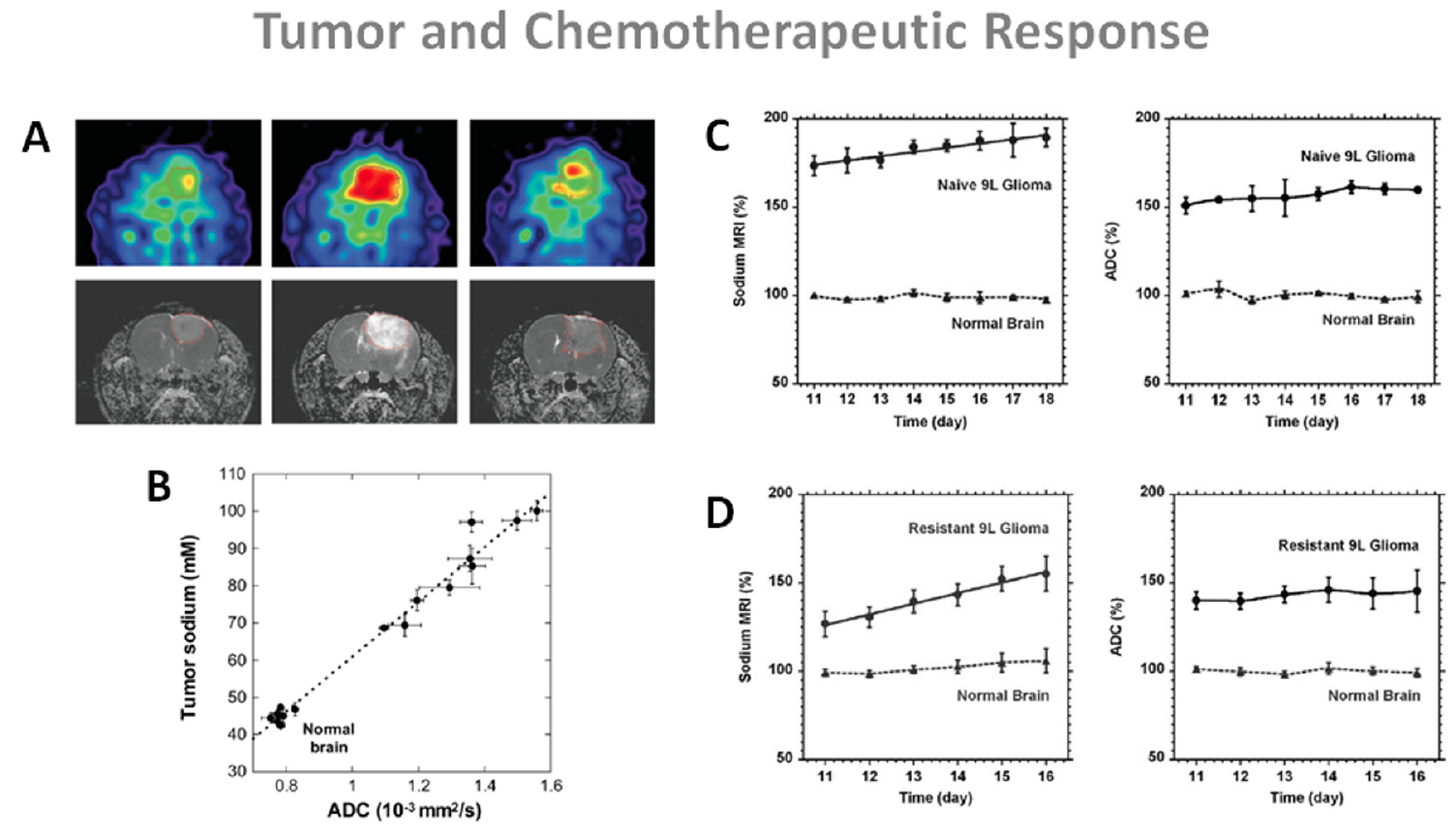}
\par\end{centering}

\caption{\label{fig:chemo}Examples of sodium MRI and proton DWI studies on
tumor and chemotherapeutic response in a rat glioma. \textbf{A}. Sodium
images (top) and proton ADC maps (bottom) of a BCNU-treated 9L rat
glioma acquired at day 0, 7 and 23 (left to right) after BCNU injection,
performed at day 17 after tumor implantation. Central sodium image
at day 7 show a dramatic increase of sodium concentration throughout
the entire tumor area. Image at day 23 shows tumor regrowth after
tumor shrinking started at day 9 and its maximum regression at day
16. \textbf{B}. Correlation of tumor sodium concentration and ADC
in the rat glioma obtained at various points following a single dose
of BCNU. \textbf{C}. Time course after tumor implantation of sodium
(left) and ADC (right) variations for a naive type 9L glioma. Sodium
and ADC data are given in percent relative to normal contralateral
brain. All data are presented as mean $\pm$ standard deviation. Sodium
concentration steadily increased in the naive tumor with a rate of
2.4\% per day while ADC was practically unchanged (1.4\% per day).
\textbf{D}. Time course after tumor implantation of sodium (left)
and ADC (right) variations for a glioma created from a resistant 9L
cell line. Sodium concentration steadily increased in the resistant
tumor with a rate of 5.8 \% per day while ADC was practically unchanged
(1.2\% per day). Sodium values were corrected for partial volume effect.
Figures A and B from Schepkin V.D et al. Magnetic Resonance in Medicine
53:85 92, 2005. Figures C and D from Schepkin V.D. et al. Magnetic
Resonance in Medicine 67:1159 1166, 2012. Reproduced by permission
of Wiley-Blackwell.}
\end{figure}

Schepkin at al. \cite{2006_19_8} studied the possibility of using
DWI and sodium MRI for assessing non-invasively the effect of chemotherapeutic
treatment of tumors \emph{in vivo}. The study was performed on 9L
gliosarcomas implanted in rats and then treated with varying doses
of BCNU. MRI were acquired at 21.1T. Main results are shown in Fig.
\ref{fig:chemo} A-B. This study demonstrate that there is a very
good correlation between DWI and sodium MRI in their ability to reveal
the efficacy of tumor therapy in a few days following treatment in
a dose dependent manner. As both these methods rely on biophysical
properties driven by changes in cellular structure (though slightly
different), it was expected that they should provide similar results. 

Babsky et al. \cite{2007_25_7} implanted fibrosarcoma (RIF-1) tumors
in mice and monitored the effects of 5-fluorouracil (5FU) treatment
with proton DWI, SQ and TQF sodium MRI and also PET for measuring
Fluorodeoxyglucose (FDG) uptake in the tumor. Tumor volumes significantly
decreased in treated animals and slightly increased in control tumors
ones 2 and 3 days post-treatment. SQ sodium signal intensity (SI)
and water ADC increased in treated tumors but not in control tumors
during the same period. TQF sodium SI and FDG uptake were significantly
lower in treated tumors compared with control tumors 3 days after
5FU treatment. They concluded that the correlated increases in SQ
sodium SI and water ADC following chemotherapy might reflect an increase
in extracellular space, while the lower TQF sodium signal and FDG
uptake in treated tumors compared with control tumors suggest a shift
in tumor metabolism from glycolysis to oxidation and/or a decrease
in cell density. A multiparametric approach using $^{\mathsf{1}}$H+$^{23}$Na
MRI, PET and histology could therefore provide a clearer understanding
of the relationship between metabolic and ionic changes in tumors
caused by chemotherapies. 

Kline et al. \cite{2000_6_6} used sodium MRI with IR to monitor the
response to chemotherapy of mouse xenograft tumors propagated from
human prostate cancer cell lines, in order to increase the weighting
of the images toward intracellular sodium content. A 37\% increase
in sodium signal intensity was observed between images before and
24h after administration of antineoplastics, which may due to an increase
of intracellular sodium in the treated tumors due. These findings
can be matched with experiments with these same drugs and cells treated
in culture, where a significant intracellular sodium elevation (10-20
mM) was detected using a ratiometric fluorescent dye. Flow cytometry
also showed that this intracellular sodium increase preceded cell
death by apoptosis. Sodium MRI with IR seems to increase the weight
of the images to intracellular sodium and could help monitoring non-invasively
apoptosis of tumors cells induced by chemotherapy.

\subsection*{Tumor Resistance to Therapy}

Schepkin et al. \cite{2012_67_4} also studied the capability of sodium
MRI and DWI for monitoring the tumor resistance to chemotherapy in
intracranial rat 9L gliomas, in order to see if these methods could
be used as biomarkers for drug resistance. It was first measured that
implanted resistant 9L cells created tumors with significantly reduced
TSC (57 mM) than nonresistant (naive) 9L cells (78 mM) and that corresponding
differences in ADC were less pronounced but still statistically significant. 

The main results are presented in Fig. \ref{fig:chemo} C-D which
shows the time course evolution of sodium content and ADC in naive
and resistant tumors after chemotherapeutic injection (BCNU). Both
Sodium and ADC can differentiate resistant from naive tumors. It is
shown that sodium content and ADC vary at very different rates in
the two kinds of tumors after treatment and thus must depends on different
mechanisms. Many parameters are involved such as tumor volume change,
increased extracellular volume, increased intracellular sodium concentration
due to deficit in ATP production, increase of blood supply, lactate
overproduction (indicator of increased glycolysis). As lactate production
tends to correlate with tumor aggressiveness, the increasing intracellular
sodium content could be an indicator of growing tumor malignancy.
No definite answer can be given but it seems that an increase in intracellular
sodium might be the main mechanism involved in the detected increase
of sodium signal.

This study shows that TSC measured with sodium MRI could give important
information about the level of drug resistance before chemotherapy
and that it is more sensitive than DWI for detecting small changes
in tumor resistance.

\chapter*{Conclusion}

\addcontentsline{toc}{chapter}{Conclusion} 

We have seen that sodium MRI can help assess directly, in a non-invasive
and quantitative manner, some important new metabolic information
such as tissue viability, through cell integity and energy status,
and that this information cannot be determined by standard anatomical
or functional proton MRI or other non-invasive imaging modality. 

From the research point of view, there is an increasing interest in
this technique, as more and more tissues and diseases are investigated,
from brain tumors or strokes to osteoarthritis in articular cartilage,
passing through kidney impairments and diabetes in muscles. However,
more research still needs to be on both software and hardware for
improving the data acquisitions and reconstructions, and increasing
the sensitivity of the technique to specific sodium parameters (e.g.
intracellular sodium or relaxation times). 

From the medical point of view, more research still has to be performed
\emph{in vivo}, on more patients with diseases (longitudinal studies)
or healthy subjects (optimization of data acquisitions and reconstruction)
in order to convince radiologist/physicians that sodium MRI could
add some crucial information for the diagnosis but also prognosis
of diseases, for the management of patients (e.g. with stroke) or
about the possible outcomes of treatments (chemotherapy resistance). 

And last but not least, from the commercial point of view, sodium
MRI cannot become a clinical tool without the help of the major scanner
vendors which should help develop multichannel-multinuclear hardware,
new multi-tuned RF coils and new UTE sequences and reconstruction
algorithms along with the availability of high magnetic fields (>3T).

\newpage{}

\begin{small}\bibliographystyle{unsrt}
\addcontentsline{toc}{chapter}{\bibname}\bibliography{Biblio_Review_Sodium_MRI_1}

\end{small}
\end{document}